%% file: acl_latex.tex
\pdfoutput=1

\documentclass[11pt]{article}

\usepackage[]{acl}
\usepackage{graphics}
\usepackage{graphicx}
\usepackage{times}
\usepackage{latexsym}
\usepackage{multirow}
\usepackage{booktabs}

\usepackage[T1]{fontenc}

\usepackage[utf8]{inputenc}
\usepackage{subcaption}
\usepackage{makecell}
\usepackage{soul}
\newcommand{\hlc}[2][yellow]{{%
    \colorlet{foo}{#1}%
    \sethlcolor{foo}\hl{#2}}%
}
\definecolor{simplification-color}{rgb}{1.0, 0.75, 0.0}
\definecolor{paraphrase-color}{rgb}{0.55, 0.71, 0.0}
\definecolor{negation-color}{rgb}{0.82, 0.1, 0.26}

\definecolor{extractive-color}{rgb}{0.82, 0.62, 0.91}
\definecolor{abstractive-color}{rgb}{0.39, 0.58, 0.93}
\definecolor{machine-generated-color}{rgb}{0.60, 0.79, 0.95}
\definecolor{human-generated-color}{rgb}{0.98, 0.86, 0.69}
\definecolor{model-color}{rgb}{0.66, 0.66, 0.66}

\usepackage{microtype}

\title{Towards Process-Oriented, Modular, and Versatile Question Generation that Meets Educational Needs}

\author{Xu Wang \quad Simin Fan \quad Jessica Houghton \quad Lu Wang \\
  Computer Science and Engineering \\
  University of Michigan \\
  Ann Arbor, MI \\
  \texttt{\{xwanghci, oliviaaa, houghj, wangluxy\}@umich.edu}}

\begin{document}
\maketitle

\input{Sections/0-abstract}
\input{Sections/1-introduction}
\input{Sections/2-relatedWorks}
\input{Sections/3-needFinding}
\input{Sections/4-findings}
\input{Sections/5-experiment}
\input{Sections/6-discussion}
\input{Sections/7-conclusion}

\section*{Acknowledgements}
This work is supported in part by National Science Foundation through grant IIS-2046016. We thank all our participants for their time.
We thank the anonymous reviewers for their valuable suggestions on various aspects of this work.

\bibliography{anthology,custom}
\bibliographystyle{acl_natbib}

\input{Sections/appendix}

\end{document}

%% file: Sections/0-abstract.tex
\begin{abstract}
NLP-powered automatic question generation (QG) techniques carry great pedagogical potential of saving educators' time and benefiting student learning. Yet, QG systems have not been widely adopted in classrooms to date. In this work, we aim to pinpoint key impediments and investigate how to improve the usability of automatic QG techniques for educational purposes by understanding how instructors construct questions and identifying touch points to enhance the underlying NLP models. We perform an in-depth need finding study with 11 instructors across 7 different universities, and summarize their thought processes and needs when creating questions. While instructors show great interests in using NLP systems to support question design, none of them has used such tools in practice. They resort to multiple sources of information, ranging from domain knowledge to students' misconceptions, all of which missing from today's QG systems. We argue that building effective human-NLP collaborative QG systems that emphasize instructor control and explainability is imperative for real-world adoption. We call for QG systems to provide process-oriented support, use modular design, and handle diverse sources of input.

\end{abstract}

%% file: Sections/1-introduction.tex
\section{Introduction}

Decades of educational research has shown the benefit of \textit{active learning} in improving students' learning outcomes where students actively engage with the materials, e.g., through question answering and problem solving \cite{crouch2001peer, deslauriers2019measuring, koedinger1997intelligent}, compared with passively receiving lectures or reading texts \cite{chi2014icap, freeman2014active}. 
However, most instructors still use traditional teaching methods, e.g., lecturing, in large-enrolment college courses \cite{henderson2007barriers, handelsman2004scientific, stains2018anatomy}, due to their limited time and resources in creating opportunities for active learning \cite{dancy2007framework, silverthorn2006s, fagen2002peer}. 
Meanwhile, there is a growing interest in leveraging NLP methods to serve educational needs. Specifically, automatic question generation (QG) is promising in enhancing students' active learning experience by creating problem-solving activities at scale~\cite{alsubait2016ontology}.

Still, the adoption of automatic QG systems in classrooms is low \cite{kurdi2020systematic, alsubait2016ontology}, mainly because those models are only suitable for specific domains, e.g., language learning, and the generated questions are often of low quality and limited in types and difficulty levels~\cite{kurdi2020systematic, alsubait2016ontology}. For QG systems to provide more meaningful support to instructors' work, we need to address at least two fundamental challenges. First, designing questions to assist learning is a highly \textit{complex, creative, and knowledge intensive} process. Typical college instructors have received years of training on the subject domain, and their question construction process is hard to be fully automated by one single model. Second, classroom is a high-stake environment, and instructors often have \textit{pre-defined goals}. As a result, they may not prefer imperfect and fallible AI models \cite{holstein2021designing}.

To fill the gap between the design of NLP-powered QG systems and the reality in classrooms, we conducted a novel \textbf{need finding study}, with two major goals: (i) To \textit{understand how instructors create questions} of high educational value, including decision making processes and information sources used throughout the procedure. (ii) To identify key touch points to \textit{improve, design, and re-create NLP systems} that instructors would find useful in their practice. 


Concretely, we interviewed and observed 11 instructors from 7 different universities as they created questions based on readings to support student learning of the content. 
We piloted the interview study protocol for several rounds, before finalizing a \textbf{two-phase protocol}. 
In the first phase, we conducted an in-depth case study of one instructor's quiz design experience for a semester-long course. 
We extracted the instructor's output and manually labeled the context of each action. We then propose a novel method: \textbf{text replay enactment}, where we replayed the instructor's process of generating the output and annotated what NLP tasks could be used to accomplish the instructor's text transformation operations. This method successfully addresses the challenge that users are unable to directly articulate what they want from AI~\cite{yang2020re}. 
Next, we summarized a list of NLP tasks that this instructor found useful, and used them as design probes in a subsequent interview study with 10 instructors. We then analyzed our data using affinity diagrams. 
On a subset of these tasks, we reported a qualitative analysis of current NLP models' performance and pointed out major issues that should be addressed to allow classroom adoption.

Our major findings include: 
1) None of the instructors interviewed are currently using any type of automatic question generation tools, though they expressed a strong desire to find better ways to design quality questions to support active learning and effective reading. 
2) Instructors' natural behaviors and thought processes reveal multiple reasons why existing automatic QG techniques fall short: They leverage multiple sources of information that cover domain knowledge, educational goals, and student misconceptions; They view question creation as an iterative process; They often apply a suite of techniques and strategies. All of these traits are largely ignored by existing end-to-end QG models.
3) Finally, instructors also reported challenges when constructing questions, and showed strong interests in receiving suggestions and support from NLP systems if the outputs are of high quality and interpretable, and that they have sufficient control when interacting with the system.


Based on these findings, we argue that developing \textit{effective human-NLP collaborative QG systems} is a promising direction, and propose \textbf{three design implications}. 
First, it is critical to provide \textit{process-oriented support} to instructors, rather than automating the process end to end.
Second, we advocate for \textit{modular system design with different NLP components} that give instructors strong control over which NLP components to use to meet their goals. 
Third, we argue that future QG systems should accept \textit{diverse data input}, including expert domain knowledge, students' background, educational goals, and pedagogical context.  

Finally, we stress the value of understanding stakeholders' needs to better design and deploy NLP systems.
As anticipated, users have concerns about how well NLP can do, and are hesitant to use AI-based tools in high-stake environments. 
Therefore, an important first step in developing human-NLP collaborative systems is to understand user needs and increase user buy-in. As we learned from the interviews, using simple models with reliable performance in the beginning while requesting more user input may be a good approach to bootstrap. We also encourage our community to collaborate with HCI researchers to develop user-friendly interfaces and investigate users' adoption and preferences of NLP tools in context of use.\footnote{Data, code, and models used in this paper are released at: \url{https://github.com/Olivia-fsm/P2MCQ}.}

%% file: Sections/2-relatedWorks.tex
\section{Related Work}
\label{sec:relatedwork}


\subsection{Automatic Question Generation (QG)}
Here we focus on describing QG models for educational purposes, and refer readers to a detailed literature review in \newcite{kurdi2020systematic}. Existing QG systems primarily produce questions that target \textit{low-level cognitive skills}, such as factual wh-questions and cloze questions that can be answered with short phrases~\cite{heilman-smith-2010-good,chali2015towards,olney2012question}. 
These simple questions only contain limited concepts~\cite{song2016question} and, as a consequence, offer limited opportunities to control question difficulty level~\cite{alsubait2016ontology} and are limited for assessment only, leaving many learning opportunities unfulfilled. Additionally, prior work largely develops domain-specific techniques for learning language, math, and medicine knowledge, relying on existing domain ontology~\cite{alsubait2016ontology, stasaski2017multiple}. 

%
%
Our study investigates multiple-choice question (MCQ) generation~\cite{ch2018automatic}. MCQs can be graded automatically and offers immediate feedback to students, which has profound benefits to scale active learning. Prior research has shown compelling results on the educational value of MCQs in comparison to open-ended question items. For example, Smith and Karpicke found that students performed equally well on English reading tasks no matter whether they practiced with multiple-choice, short-answer, or hybrid questions \cite{smith2014retrieval}. Similarly, multiple-choice questions are shown to provide a win-win situation compared to open-ended cued-recall tests on English reading tasks \cite{little2015optimizing,little2012multiple}. 
The authors find that both open-ended and cued-recall tests foster retention of previously tested information, but multiple-choice tests also facilitated recall of information pertaining to incorrect alternatives, whereas cued-recall tests did not. Recent work \cite{wang2021seeing} demonstrates that even for less well-defined domains, MCQs could exercise critical thinking elements and require students to evaluate the quality of MCQ options, especially when common student misconceptions were used as distractors \cite{wang2019upgrade}.


This poses unique challenges in automatic generation of MCQs. First, creating rich-content and meaningful correct options that help students retrieve correct and relevant information. Second, generating plausible distractors that help students internalize incorrect information. Well-designed and meaningful distractors require students to evaluate and contrast options \cite{wang2019upgrade}, that involves going beyond surface traits and considering deeper connecting principles \cite{schwartz2011practicing}.
However, only phrase-level replacement has been commonly studied~\cite{papasalouros2008automatic} in automatic generation of MCQ options and distractors. Recent automatic QG systems are built on end-to-end trained neural generation models, where a single question is generated from a given context~\cite{sun-etal-2018-answer,zhou-etal-2019-question,zhang-bansal-2019-addressing}. 
Although questions that require multi-hop reasoning~\cite{pan-etal-2020-semantic,su-etal-2020-multi} or reading long text~\cite{bi-etal-2021-simple-complex,cheng-etal-2021-guiding,cao-wang-2021-controllable} have been studied, these systems do not offer control over question difficulty nor consider different sources of knowledge. 
In this work, we propose to modularize the automatic QG process with different components, e.g., summarization, simplification, or contradiction generation, to offer flexible interface for instructors to control various aspects of the produced questions.

\input{Sections/2-3-NLP-subtasks}
\input{Sections/2-4-human-machine-collaboration}

%% file: Sections/2-3-NLP-subtasks.tex
\subsection{NLP Tasks for Education} 

NLP systems have been developed to support a variety of educational applications, including writing assistance~\cite{Bellino2020DesignAE, FrankenbergGarcia2018DevelopingAW}, reading comprehension support ~\cite{Ross1991SimplificationOE, Vodolazova2019TowardsAT, Vajjala2019OnUT, siddharthan-katsos-2010-reformulating, Chatzipanagiotidis2021BroadLC}, language learning systems~\cite{Tweissi1998TheEO, Petersen2007TextSF, Katinskaia2021AssessingGC, ksik2021EstonianAA}, and generating feedback for programming and design~\cite{Wang2018SearchAA, Kang2019AutomatedFG}. 
Here we provide a brief overview of the state-of-the-art NLP models that are relevant to the question generation process. 

\textbf{Summarization}, the ability of condensing a reading material into a concise passage, has been used for evaluating and improving students' reading comprehension ability~\cite{edmonds2009synthesis,vaughn2011efficacy,stevens2019review,Hwang2019EffectsOI}. 
Similarly, \textbf{paraphrasing} also demonstrates students' content comprehension skills~\cite{Haynes1984ParaphrasingAR} since it requires the ability of conveying the same semantic meaning with different languages. However, both tasks are rarely studied for question generation~\cite{lyu-etal-2021-improving}. \textbf{Simplification}, on the other hand, has demonstrated its values in language learning and reading comprehension~\cite{Tweissi1998TheEO,inui2003text,Petersen2007TextSF,Rets2021ToSO}, and is often treated as a preprocessing step to convert complex sentences into simpler versions before creating questions~\cite{majumder2015system,patra2019hybrid}. 
State-of-the-art NLP models for these three tasks are all based on neural generation systems, which are known to suffer from errors and lack of controllability~\cite{maynez2020faithfulness}. Therefore, their usefulness for QG will largely depend on output quality and new designs with enhanced explainability and easy-to-use interface.

%% file: Sections/2-4-human-machine-collaboration.tex
\subsection{Human-AI Systems for Education}
The concept of human-AI systems is recently introduced into the education domain, where human inputs are continuously solicited and there exists a collaborative relationship between humans (often teachers) and AI algorithms to provide effective instruction to students. They differ from prior AI for education systems where human (teacher and student) input is often elicited before the development of the system \cite{koedinger1997intelligent, kumar2007tutorial}. 
Human-AI systems have been developed to help human teachers more easily identify the students that are struggling \cite{holstein2018student}, design higher quality assignment questions \cite{wang2019upgrade}, and offer aggregated feedback to students' programs \cite{glassman2015overcode}. They all demonstrate the advantages of combining both humans (robust and flexible) and AI (low-cost and quick) in addressing real-world challenges in education, and it inspires us to develop human-AI systems for QG.

Also relevant to this work is the abundant literature on human-AI interaction \cite{horvitz1999principles, Amershi2019GuidelinesFH, yang2020re}, which highlights the need for considering user's goals and behaviors and points to general design guidelines when prototyping and designing algorithmic experiences. We consider this work to be an instantiation of the idea in a specific context, examining human-NLP interaction when instructors design questions for educational purposes. We expect our findings to contribute to the future development of human-AI systems that address this specific educational problem, and to the greater body of work on human-AI interaction and design principles.

%% file: Sections/3-needFinding.tex
\section{Need Finding Study Methodology}
\label{sec:needfinding}

We wanted to understand how instructors construct questions that align with their educational needs. We also wanted to probe on when instructors think an intelligent system might offer support or their work. To address these needs, we chose to conduct a qualitative study consisting of observations and semi-structured interviews. It has become a standard HCI approach when designing new software systems that address human needs. 

In this work, we conducted an IRB-approved two-phase need finding study. We went through rounds of piloting before reaching at the final protocol, which addressed two challenges that emerged during the pilots. 1) An interview-alone approach is insufficient since it is hard for instructors to recollect all details of question design. In addition, it requires concrete textual input for us to understand how humans use texts when designing questions.  
2) Instructors are not able to directly articulate their NLP needs \cite{yang2020re}. 
To address the first challenge, we propose a specific scenario where instructors design multiple-choice style quiz questions to support active reading. On one hand, it targets an authentic problem since text reading is a passive learning experience that occurs everyday in college classrooms. Active reading opportunities are much needed to support students' comprehension and learning. 
On the other hand, 
it is a question design scenario where the text input is explicit. 
To address the second challenge, we propose a text replay enactment method where we, as NLP researchers, review users' operation on the text and annotate possible NLP tasks that can be applied to make the text transformation.






\subsection{Phase 1: Case Study and Replay Enactment of Text}

The case study contains an in-depth analysis of one instructor's quiz design experience for a semester-long course on human-computer interaction at the University of Michigan. We will refer to the course as HCI101 for the rest of this paper. The course has 1-2 required readings per lecture, including academic papers and book chapters. The instructor (Instructor A) designed quiz questions for each reading text to facilitate students' reading and understanding of the material.

\begin{figure*}[htbp]
    \centering
    \includegraphics[width=\textwidth]{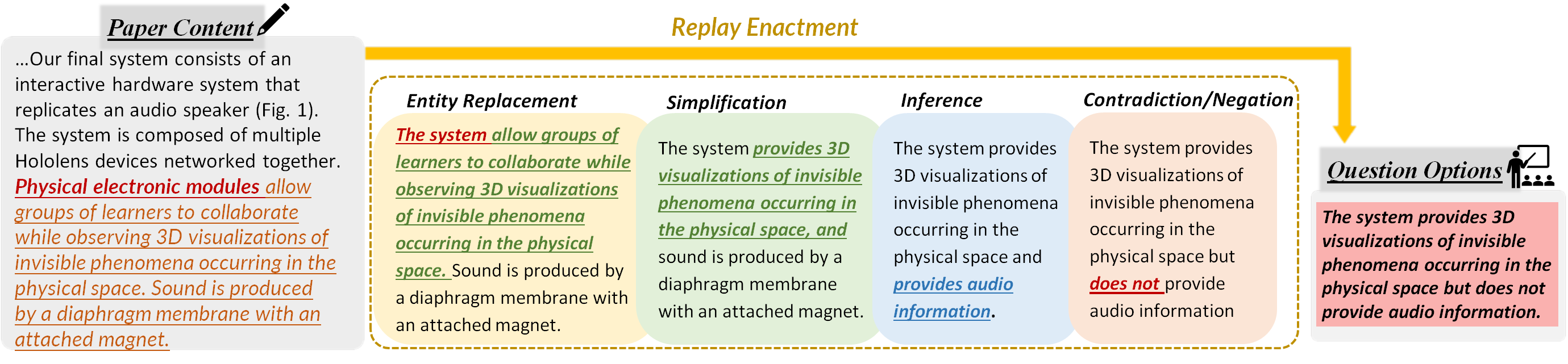}
    \caption{An example Text Replay Enactment process. For every context-option pair, we as NLP researchers annotate what NLP modules are needed to generate the final question option from the original context. In this case, 5 NLP modules: Extraction, Entity Replacement, Simplification, Inference, and Negation are needed.
    }
    \label{fig:flow-diagram}

\end{figure*}


\subsubsection{P2MCQ Dataset}

We obtained all questions that Instructor A created during HCI101, including 160 multiple-choice questions with 629 question options in total (197 correct answers and 432 incorrect answers or distractors).
These quizzes were manually written based on 30 reading materials (24 conference papers and 6 book chapters) and were assigned throughout HCI101. Example questions can be seen in Figure~\ref{fig:MCQ-sample}.

\paragraph{Context Annotation}
One annotator aligned each question option with its supporting context in the original reading material. 
Both sentence-level and paragraph-level contexts are extracted. For each single option, the sentence-level context include sentences with its supportive evidence, and the paragraph-level context refers to the whole paragraph containing those sentences.
During context annotation, questions and options without supportive evidence in the given material (i.e., they were designed based on the instructor's prior knowledge, not from the text) were removed.
Instructor A also checked our annotations for accuracy. This resulted in a dataset of 128 multiple-choice questions, including 439 question options (110 target options and 329 distractors). We release the P2MCQ dataset with this paper.

\subsubsection{Text Replay Enactment}

We are inspired by user-centered design methods, including user enactment \cite{odom2012fieldwork} and replay enactment \cite{holstein2020replay}, through which designers imagine possible futures and use them as conversation starters to elicit users' feedback on technologies that do not exist in users' previous experience. We propose a text replay enactment method where researchers analyze user's written text and the context, to annotate what possible NLP tasks could help the user achieve this outcome. A sample enactment process is shown in Fig~\ref{fig:flow-diagram}.
To ensure that the annotation process is consistent with the instructor's intention, we started with a reflective interview with Instructor A on a sample of the P2MCQ dataset (10 question stems and 40 options).
Instructor A described how they arrived at the question stem and each of the options.
We then applied the text replay enactment process on the entire P2MCQ dataset. 
All four researchers first constructed the annotation scheme collaboratively with 5\% of the dataset. Two researchers then completed another 13\% of the dataset and verified inter-rater reliability. One researcher further completed the annotation. The resulted annotation scheme with NLP tasks and examples can be found in Table~\ref{table:2}. These NLP tasks are used in phase 2 to probe into users' needs for NLP support.

\subsection{Phase 2: Interview and Observation Study}
We recruited participants through social media (including mailing lists and social groups of professors) and offline correspondences.
10 instructors from 7 different universities participated in the study. The instructors have teaching experience ranging from 2 to 40 years and are from disciplines including computer science, information science, data science, education, developmental psychology, and political science.

The interviews were done through Zoom and each lasted between 50 and 75 minutes. Participants were given a \$50 Gift Card. We first asked participants to share how they approached reading assignments. The majority of the session was spent having the participant design questions based on a reading text of their choice. Specifically, we asked them to design questions (MCQ preferred) that could help their students understand and learn from the content. 
We asked the participants to think aloud throughout the process.
We provided question templates sampled from the P2MCQ dataset to help them jump start. 
Participants were able to design 3 to 10 multiple-choice questions (with one question stem and four options) during the session.
We then asked the participant to reflect on how they arrived at each question stem and option, and shared the challenges they had encountered throughout the process.
At last, the researcher asked the participant to imagine there being an intelligent system to provide support alongside the quiz design process. For each of the NLP tasks in Table \ref{table:2}, we presented the task in the context of the user's own quiz design experience, and asked the participant to share to what extent did they find the support useful. We transcribed the interview recordings and analyzed our data using affinity diagrams \cite{moggridge2007designing}, which is a commonly used method in HCI to identify emerging themes from qualitative data and discover opportunities for technologies to enhance future.

%% file: Sections/4-findings.tex
\section{Findings}
\label{sec:findings}

Through affinity diagramming, we observe emerging themes from the interview data. We present our findings and use participants' quotes to illustrate these themes. We refer to participants as P1-P10.


\subsection{Supporting active reading is desirable yet expensive}
All participants mentioned that there should be better ways to support students in reading. For example, P5 said ``\textit{It is definitely a problem that instructors face}.''
Participants shared the techniques they have used to support reading and the limitations of such approaches. For example, reading summaries is not scalable as grading can be challenging; collaborative annotation platforms such as Perusall \cite{Perusall} encourage participation but do not necessarily make sure students get the key messages.
Most participants expressed interest in using this approach if question design becomes less expensive and more accessible, as P2 said ``\textit{I wouldn't have the time to do it this way, but if you were to give me the questions, I think it would be amazing.}''


\subsection{Why Automatic Question Generation Techniques Are Insufficient}
Instructors often use the questions for specific educational goals, leverage diverse data sources when designing questions, and iteratively improve the content.
We point out five categories of reasons on why existing QG techniques are insufficient, as they overlook these critical factors behind making questions of high educational value.

\subsubsection{I want the questions to be aligned with my teaching goals}
\label{sec:goals}
Instructors are more interested in targeting higher-level concepts instead of specific details, when supporting active reading. P2 mentioned ``\textit{students complain when asking specific questions about the reading, because they feel like we're not testing them for knowledge, we're testing them for obscure pieces of information in a paper.}'' Other instructors also worry that using detail-oriented quiz questions may make the students perceive it as a reading comprehension task and mindlessly answer the questions without thinking about the broader implications of the reading. 
Instructors require the quiz questions to meet their objectives of assigning the reading and help students ``get a gist'' and ``think deeply'' of the material. For example, P10 mentioned ``\textit{I'm never trying to just see how much they can memorize. We are really trying to get at the conceptual stuff.}'' 
P1 wanted the students to think critically about the assigned text: ``\textit{This paper itself has all of these pretty deep flaws, and I want students to see the flaws.}''


\subsubsection{This is not based on the text. It's from my own experience.}

An emerging theme from the interviews is that instructors rely on their prior knowledge when designing questions that they think are of educational value. This is a frequent quote we get from participants ``\textit{So this [question option] I'm not getting from the article. This is from my own experience.}''
There are three categories of information that instructors leverage: 1) key messages in the text; 2) common student misconceptions; 3) course syllabus and prerequisite relationships. 

First, instructors rely on their memory of what are \textbf{the key takeaways from the text}. They would often skim the reading text and say ``\textit{I want people to get right away [a concept]}.''

Second, instructors use \textbf{students' common misconceptions} when designing questions and distractors. 
For example, P5 said ``\textit{Most students got it wrong, because they have a shallow understanding of the content.}'' 
P10 also made similar comments ``\textit{Students have trouble distinguishing what's a milestone and what's a mechanism, so I could imagine a question\ldots.}''

Third, instructors consider \textbf{syllabus information} and think about how the questions are connected to previous or subsequent activities in the course. 
For example, P2 said ``\textit{[I'm interested in] quiz questions that are heavily integrating a broad set of concepts throughout the course. I want to test them on the whole understanding of the material\ldots}.''



\subsubsection{The distribution and sequence of questions are also important}
In addition to the questions themselves, instructors also keep thinking about the meta-level properties of the quiz, such as the distribution and sequence of questions. E.g., P4 said ``\textit{I want to make sure my questions cover all three sections in that chapter.}'' P6 paid attention to the sequence of questions as they were in particular interested in using the questions as a reading guide. 



\subsubsection{It is an iterative process}
Another frequent quote from the quiz design processes is that ``\textit{I'll write it down first and see whether I like it.}'' Multiple participants mentioned that ``\textit{this may not be the final question}'' or ``\textit{I'd like to make this better.}'' Question design is an iterative process where instructors keep revising the text and may jump between questions. 





\subsubsection{Instructors combine a suite of techniques and strategies}
One emerging theme from both phases of our study is that instructors combine a variety of strategies when designing questions and apply different strategies across contexts. From text replay enactment, we found that multiple text operations are often needed (in 57\% of the dataset) in order to produce the target text used in the question (Figures~\ref{fig:task-breakdown} and \ref{fig:task-numbers}). 




\subsection{Challenges in Human QG}

The two mostly mentioned challenges are 1) \textbf{figuring out the key messages} that instructors want students to take away from the text; 2) \textbf{coming up with distractors} (incorrect options). Multiple participants described how identifying the key messages is critical and hard. P4 expressed ``\textit{I think the challenging thing is to find out the outcome I want students to have after reading the article}.'' P5 also mentioned ``\textit{I think one challenge is like, I realized there's a lot of information in the reading, and seeing what are the key things I want students to get across}.''
Additionally, all participants found coming up with plausible distractors to be difficult. P5 commented ``\textit{coming up with the distractors, it is challenging. Because I don't know what would distract the student, but also not be too confusing.}'' 

For novice instructors, \textbf{producing the question stem} is difficult.
We saw participants often referred to the question templates for ideas.
Participants also requested \textbf{meta-level monitoring}, e.g., making sure the difficulty of the questions is reasonable, the questions have a good coverage of the content, etc.
Some instructors mentioned it was challenging to \textbf{ensure that questions are ``good''}. P7 expressed their frustration that ``\textit{What I always want, and never had in a multiple-choice quizzing tool is a way to say if they've chosen this answer, they have demonstrated these learning objectives and failed to demonstrate these learning objectives.}'' 
P5 said ``\textit{maybe I write a question, and then there could be support like `it could have been framed better.' Bringing in learning sciences principles and suggest how to write good questions would be useful.}''
Finally, some encounter challenges in \textbf{phrasing and wording}. P6 mentioned, ``\textit{the hardest part for me is to decide how to rephrase it.}'' 


\subsection{Desires for NLP Support}
In general, participants showed positivity towards receiving support in the process, and would want to spend the minimal time possible on question design.  
Participants had mixed opinions about the proposed NLP support concerning the level of control they could have and the performance of the NLP models. Instructors considered some models to be beneficial in some cases but not all. We also observed individual differences on which models they found useful. 
For example, for summarization, P7 did not think it would be useful since the reading text was already very condensed, but P6 and P9 felt that it would help shorten their own rereading time and extract useful information.
Participants consider many tasks to be really hard to be automated, and questioned how the models would be able to produce the desirable outcomes. 
For example, P7 felt that in order to use simplification, the model performance would have to be perfect, because they are concerned with losing important nuances or misrepresenting ideas in the actual text. Ultimately, the participants were interested in trying the system, and thought that it could help them perform QG faster.  

%% file: Sections/5-experiment.tex
\section{Evaluation of Existing NLP Tools}
\label{sec:nlp}

Given the observations from \S\ref{sec:findings}, here we select popular NLP models for the most frequent text transformation operations used by the instructors and investigate existing issues. We show both instructor-constructed transformations and system outputs in Table~\ref{tab：nlp-sample} 
in Appendix~\ref{appendix:B-sample}. 
Model implementation and dataset collection details are given in Appendix~\ref{sec:appendix-A}. 

Here we use the context consisting of one or multiple sentences annotated on the quiz dataset as described in \S\ref{sec:needfinding}. 
We first conduct \textbf{sentence selection} 
and \textbf{abstractive summarization} based on \textsc{\textbf{BertSumExt}} and \textsc{\textbf{BertSumExtAbs}} models~\cite{Liu2019TextSW}, which are extractive and abstractive summarization models fine-tuned on news articles. 

To build an abstractive summarization model that is suitable for scientific domains, we further fine-tune the sequence-to-sequence BART model ~\cite{lewis-etal-2020-bart} on our newly collected HCI article summarization dataset that consists of section-summary sentence pairs. 

Given the same context, we then build models for text \textbf{simplification} using \textsc{\textbf{ACCESS}}~\cite{Martin2020ControllableSS} and \textsc{\textbf{MUSS}}~\cite{martin2021muss}, which are built on top of BERT and BART, respectively; a \textbf{paraphrasing} model that fine-tunes BART using \emph{ParaSCI}~\cite{Dong2021ParaSCIAL} which contains paraphrase pairs from scientific papers; a \textbf{negation} generation model based on \textsc{\textbf{CrossAUG}}~\cite{crossaug2021} which fine-tunes BART to produce text that contradicts the given context. 


Three major findings are made. For sentence selection and summarization models, it is often hard to interpret why the output is produced as is. For example, it is unclear why a sentence is treated as more important and thus selected from a given context. 
Second, for the generation-driven models, such as abstractive summarization, simplification, and paraphrasing, their output is sometimes only tangentially relevant to the content. The output frequently contains content that cannot be inferred from the input as well as errors, which can hurt instructor experience and even raise ethical concerns. 
Moreover, the diversity of the generations is rather limited. For example, simplification and negation models tend to focus on changing specific words. 
Overall, all these NLP models are trained without user needs being specified, thus do not offer control over which topic to summarize, what phrase or content to simplify or paraphrase, and which knowledge to be used as a pivot to create distractors. 
These issues point to the future directions for building summarization models with \textit{explainability}, \textit{guaranteed faithfulness and correctness}, as well as that \textit{allow users to exert control} over where the transformation should be applied.

%% file: Sections/6-discussion.tex
\section{Implications for Research}
\label{sec:discussion}


Our study reveals instructors' natural processes of constructing questions, the challenges they have, and when they may benefit from NLP support. Specifically, we see a strong desire for user control, where humans provide input to NLP systems and can decide when to use NLP outcomes. 

\subsection{Implications for Developing Human-NLP Collaborative QG Systems}

\textbf{Recommendation 1: Instead of generating outcomes, providing process-oriented support is more desirable.}
First, instructors considered it to be highly critical for the questions to serve their teaching goals. 
As P10 put it ``\textit{Whatever the goals are, they are shaping those questions, because mostly, the texts, chapters or articles, they were not written with the course in mind. I have to take a reading that was meant for one purpose and pull it to my purpose.}'' 
When QG systems do not align with instructors' goals, it's hard for them to meet the educational needs and support student learning.
Second, all participants viewed question design as an iterative process and preferred to have the opportunity to keep revising and improving content. 
Third, we observed that instructors had challenges in doing QG themselves, e.g., when the text was dense, refreshing their memories of the key messages was time-consuming; when they wanted to include a distractor on one concept, figuring out the exact language for an option was hard; instructors may also jump around the whole article and make inferences based on text spread in the article.   

We argue that it is more productive for NLP systems to provide process-oriented support to instructors instead of trying to generate complete outcomes immediately. As an example, most instructors liked the idea of highlighting key phrases in the text for the system to identify relevant content. Instructors also liked having the system summarize content for them to decide what questions to design. 
When asked about using Contradiction/Negation to create distractors, they applauded the idea of telling the system which parts to negate. We see a future direction of building human-NLP collaborative systems where expert input is continuously solicited to tell the system where to pay attention to. 




\textbf{Recommendation 2: Develop QG systems with NLP modules that provide the flexibility of applying modules depending on the context.} 
Instructors had different preferences on which NLP modules to use depending on the goal of the questions, students' prior knowledge, etc. For example, instructors want to use easy questions at times as quick comprehension checks: ``\textit{This one [question] is actually quite easy. If they read the paper they'll know.}'' Other times, they want to make challenging questions that provoke deep thinking, as we discussed in depth in \S\ref{sec:goals}. 
Even for a single question, instructors prefer using different strategies to design options. E.g., P5 suggested using Contradiction/Negation to generate one distractor and using Extracting Parallel Concepts to generate another. This ensures that a question contains rich information that benefits student learning. Additionally, we also observed substantive cases where instructors chained multiple NLP modules (as shown in Figure~\ref{fig:flow-diagram}) to generate an option.
%

We argue that framing QG as a single NLP task does not align with educators' process of question creation. Instead, we encourage QG systems to provide modular NLP tools and give instructors the flexibility of choosing which NLP tools to use as they see fit.
This approach will also greatly increase instructors' trust to the system and the interpretability of QG systems. We provide a list of NLP tasks that instructors find useful (Table~\ref{table:2}), the frequency of these tasks in the P2MCQ dataset (Figure~\ref{fig:task-breakdown}), and the preliminary results on the performance of existing NLP tools on these tasks (Table~\ref{table:1}).




\textbf{Recommendation 3: QG systems need diverse data sources as input.}
Traditional QG systems mostly use only one text source as input. We observed instructors relying on diverse data sources when designing questions. Some instructors wanted to use student-created examples in questions to reflect common misconceptions. P10 said ``\textit{Since I have lots of examples of what students have said, I could use student examples and say, which ones of these are good examples?}'' 
P7 mentioned an interest in using the whole course content as a data source: ``\textit{I could imagine a system taking not only the paper in but also taking in the rest of my course content, just as secondary, like what is foreshadowing, what is he going to teach in about two weeks.}'' Even within a single document, we observed that users may aggregate texts from non-adjacent locations as input.
We encourage QG systems to take diverse data sources as input, including previous student answers, course syllabus, lecture notes, relevant reading materials, and give instructors the flexibility to select the input sources.





\subsection{Implications for Research on Human-Centered NLP}

\subsubsection{Towards More Robust NLP Outcomes}
As detailed in \S\ref{sec:nlp}, we observed a considerable quality gap between human-generated options and machine-generated ones.  
First, end-to-end neural models often produce extraneous information. 
To address this, we propose to solicit user input and train models to focus on user-specified requirements. 
As an example, for sentence simplification, user may choose to keep several phrases and ask the system to simplify the rest of the content. We encourage researchers to work on models that support and comply with different forms of user control. 
%
Second, we argue for modular NLP systems. With the large language models becoming the de facto tool for NLP tasks of varying difficulty levels, dividing the tasks into steps and chaining the outputs could improve the quality of task outcomes~\cite{wu2021ai}.
Our study revealed that instructors applied complex transformations on text when creating meaningful questions. We further proposed a list of pre-defined NLP tasks that instructors frequently employed, and evaluated fine-tuned language models for these tasks. In future work, we consider the inclusion of user input as part of the modular system design, e.g., allowing users to construct prompts or instructions~\cite{floridi2020gpt,wei2021finetuned} in-situ to interact with the large language models to better satisfy their needs.
%

\subsubsection{Prototyping Human-Centered NLP Systems}
In this study, we also surface challenges in designing and developing human-centered NLP systems.
First, users are not able to articulate their needs around NLP. For example, users are not able to say they want the system to make an inference or complete an entity replacement at a certain point. 
We consider the text replay enactment approach to be effective in helping researchers understand user intentions and desires for NLP and encourage others to further explore and extend this method.
Second, we found that having users directly evaluate NLP outcomes is hard, as it often requires them to read the context and outputs by several models.
When we tried this in a pilot interview, a user found this to be too cognitively demanding. A more effective way is to visualize the NLP outcomes in context to reduce cognitive load, and investigate user adoption and preferences in real usage. We also encourage collaboration between NLP and HCI researchers in making progress on novel methods to prototype NLP experiences and build usable NLP systems.





%% file: Sections/7-conclusion.tex
\section{Conclusion}
\label{sec:conclusion}
QG has been an area of interest in the NLP community. However, the adoption of QG systems in classrooms is low. The goal of this work is to investigate this gap and explore directions for future QG systems that can meet stakeholders' needs.  
Our work reveals that existing QG systems do not take information that instructors deem critical when designing questions to support learning, including educational goals, and student misconceptions. We surface instructors' desires for receiving support during question design. We make recommendations on how to develop process-oriented, modular, human-NLP collaborative QG systems. 


%% file: Sections/appendix.tex
\appendix
\newpage
\section{Details for NLP Models on Selected Tasks}
\label{sec:appendix-A}
Here we describe implementation details for the NLP models used in \S\ref{sec:nlp}. 
\subsection{Sentence Extraction}
We extract salient sentences from the given context using extractive summarization model. 
The automatic extractive summarization task is approached as sentence classification, where each sentence $i$ is assigned a label $y_i \in \{0, 1\}$, with $1$ indicating the sentence is predicted for inclusion in the summary, otherwise $0$. 

We use \textsc{\textbf{BertSumExt}}~\cite{Liu2019TextSW} that adds a sentence positional encoding schema to the large pre-trained BERT model. In our experiment, we take the released checkpoint trained on CNN/DailyMail dataset and only consider paragraph-level context as the input. 




\subsection{Abstractive Summarization}
\subsubsection{Pre-trained Summarization Model}
We first use \textsc{{BertSumExtAbs}}~\cite{Liu2019TextSW}, which is first fine-tuned on the extractive summarization task and then on abstractive summarization. We again use a checkpoint of \textsc{\textbf{BertSumExtAbs}} fine-tuned on {CNN-DailyMail} released by the authors.



\subsubsection{Summarization Model for Scientific Domain}
To build an abstractive summarization model that is suitable for scientific papers, we build \textsc{\textbf{BartSum-HCI}} by fine-tuning BART~\cite{lewis-etal-2020-bart} on an automatically aligned section-summary pairs collected from HCI papers on \emph{arXiv}.

\paragraph{Dataset Collection (\emph{arXiv-HCI}).}

We first retrieve papers from \url{arXiv.org} with search query \emph{cat:cs.HC} to identify all HCI relevant papers, resulting in 8658 academic papers. 
Since abstract is expected to condense the most salient information of a paper, we consider each sentence in abstract as a sentence-level summary of relevant sections in the paper. 
To align each abstract sentence with the corresponding section context, we apply BM25 to rank all sections by their lexicon similarity to the sentence and picked the top 2 sections to create two section-sentence pairs. 
This creates the \emph{arXiv-HCI} dataset, which contains 72755 section-sentence pairs. 
The average numbers of tokens in context and summary are 468.62 and 23.01, respectively.

\paragraph{Continued Training}
We then fine-tune BART on \emph{arXiv-HCI} with train/validation/test splits of 71301/727/727. 
We use the checkpoint \emph{bart-base-finetuned-arxiv} released by HuggingFace\footnote{\url{https://huggingface.co/mse30/bart-base-finetuned-arxiv}}, which has been trained on scientific papers. 

The model is trained for 3 epochs, using Adam optimizer with default parameters ($\beta_1, \beta_2$)=(0.9, 0.999) and $\epsilon$=1e-08. 
The learning rate is initialized as $3\times 10^{-5}$ with 2000 warm-up steps and the weight decay is set to 0.01.
After training for 25000 steps with batch size 8, the model is evaluated with ROUGE-1 of 20.51, ROUGE-2 of 5.30, and ROUGE-L of 16.73 on the test set.

\subsection{Simplification}
We use the AudienCe-CEntric Sentence Simplification (\textsc{\textbf{ACCESS}}) model~\cite{Martin2020ControllableSS} for sentence simplification. In our experiments, we take the released checkpoint\footnote{\url{http://dl.fbaipublicfiles.com/access/best_model.tar.gz}} of \textsc{{ACCESS}}, which is pre-trained and evaluated on WikiLarge dataset. During our inference process, the default control parameters by \textsc{{ACCESS}} are used.

We also experiment with the Multilingual Unsupervised Sentence Simplification (\textsc{\textbf{MUSS}}) model~\cite{martin2021muss}. 
The dataset used for \textsc{{MUSS}} training is a combination of WikiLarge dataset and mined paraphrases from {CCNet}.

\subsection{Paraphrasing}
We create a paraphrase generation model \textsc{\textbf{Bart-Para-SCI}} by fine-tuning the \emph{bart-paraphrase}\footnote{\url{https://huggingface.co/eugenesiow/bart-paraphrase}} checkpoint on the {ParaSCI-ACL}~\cite{Dong2021ParaSCIAL} dataset, which contains 33,981 paraphrase pairs from articles published in ACL conferences and workshops. 
The model is trained for 10 epochs, using Adam optimizer with default parameters ($\beta_1, \beta_2$)=(0.9, 0.999) and $\epsilon$=1e-08. The learning rate is initialized as $3\times 10^{-5}$ with 2500 warm-up steps. The weight decay is set to 0.01.

\subsection{Negation}
To generate distractors, we use \textsc{\textbf{CrossAUG}}, which is proposed as a data augmentation method by training BART to generate negative claims~\cite{crossaug2021}. It is fine-tuned on the WikiFactCheck-English dataset, with positive claims as the inputs and their corresponding negative claims as the outputs. 
We use the checkpoint\footnote{\url{https://huggingface.co/minwhoo/bart-base-negative-claim-generation}} released by its author. 

\section{MCQ Samples with Human-generated and Machine-generated MCQ Options}
\label{appendix:B-sample}

\subsection{Context-option Generation Comparison Samples}
Table~\ref{tab：nlp-sample} lists sample machine-generated options along with human-constructed transformed texts, based on the same context from our \emph{P2MCQ} dataset (\S\ref{sec:needfinding}). 
For \hlc[simplification-color]{\textbf{Simplification}}, \hlc[paraphrase-color]{\textbf{Paraphrasing}}, \hlc[negation-color]{\textbf{Negation}} and \hlc[abstractive-color]{\textbf{Abstractive Summarization}} tasks, the sentence-level contexts are used as inputs; for the \hlc[extractive-color]{\textbf{Extraction}} task, the paragraph-level contexts are used.

\begin{table*}[htbp]
\footnotesize
\centering
\begin{tabular}{|m{5cm}|m{5cm}|m{5cm}|}
\hline
\textbf{Context} & \textbf{Human Generated Option} & \textbf{Machine Modification}\\
\hline
Other studies have found that helping people to think about themselves as having multiple identities, in particular, focusing on those facets of their identity that are in-group (e.g., college student) rather than outgroup (e.g., female), improves performance for those at risk of stereotype threat. 
& [\hlc[simplification-color]{\textbf{Simplification}}] Studies have found that helping people to think about themselves as having multiple identities could improve performance for those at risk of stereotype threat 
& \hlc[model-color]{\textsc{\textbf{ACCESS}}}
Other studies have found that helping people to think about themselves as
having many identities, in particular, using those facets of their identity that are in-group (e.g. college student), rather than outgroup (e. g , female ) , improves performance for those at risk of stereotype threat .

\hlc[model-color]{\textsc{\textbf{MUSS}}}
Other studies have found that helping people to think about themselves as having multiple identities, in particular, focusing on those facets of their identity that are in-group (e.g., female) helps them to remember who they are, even if they do not feel like they are.
\\
\hline
In the neoclassical theories and Soar, once knowledge leaves working memory it is permanently lost, whereas in ACT the knowledge remains but is inactive. 
& [\hlc[simplification-color]{\textbf{Simplification}}] Once declarative knowledge leaves working memory, it's permanently lost.
& \hlc[model-color]{\textsc{\textbf{ACCESS}}}
In this case, theories and Soar, once knowledge leaves working memory it is lost, whereas in ACT the knowledge remains but is still inactive.

\hlc[model-color]{\textsc{\textbf{MUSS}}}
In the neoclassical theory, once knowledge leaves working memory, it is lost forever. In the ACT theory, the knowledge remains but is not used.
\\
\hline
Cognitive skills are realized by production rules.
& [\hlc[paraphrase-color]{\textbf{Paraphrasing}}] The skill itself should be modeled as a set of production rules.
& \hlc[model-color]{\textsc{\textbf{Bart-Para-SCI}}}  the cognitive skills can be found by production rules.
\\
\hline
People must evaluate and improve their practice by seeing the effects of their efforts in the world.
& [\hlc[paraphrase-color]{\textbf{Paraphrasing}}] Learners need to receive in-time feedback form them to evaluate and improve their practice.
& \hlc[model-color]{\textsc{\textbf{Bart-Para-SCI}}}  they must evaluate and improve their practice by the effects of their efforts in the world.
\\
\hline
Increasing the number of raters per submission helps accuracy but quickly yields diminishing returns'
& [\hlc[negation-color]{\textbf{Negation}}] Increasing the number of raters per submission would always help increase accuracy of median grades.
& \hlc[model-color]{\textsc{\textbf{CrossAug}}} Increasing the number of raters per submission decreases accuracy and quickly yields diminishing returns
\\
\hline
The second finding is that it is not just the quantity of practice but also the quality. Chess duffers get good by playing. Grand masters get great by further studying famous games. Deliberate practice involves focusing on what is beyond one’s current skill set rather than just executing what one is already able to do.
& [\hlc[extractive-color]{\textbf{Extraction}}] Deliberate practice involves focusing on what is beyond one's current skill set rather than just executing what one is already able to do.
&  
\hlc[model-color]{\textsc{\textbf{BertSumExt}}} The second finding is that it is not just the quantity of practice but also the quality.
\\
\hline
These phenomena interact with each other in complex ways while being invisible to the naked eye, thus making the concept difficult to understand.  Yet these phenomena are critically important for understanding the physics of electromagnetism.
& [\hlc[abstractive-color]{\textbf{Abstractive Summarization}}] Electromagnetism is a complex mechanism, and many physical phenomena such as electric current are invisible.
& \hlc[model-color]{\textsc{\textbf{BertSumExtAbs(CnnDm)}}}
These phenomena interact with each other in complex ways while being invisible to the naked eye.

\hlc[model-color]{\textsc{\textbf{BertSumExtAbs(XSum)}}}
one of the world 's most important phenomena of the visual effects of last year has been explained by the un 's visual assessment of the effects of electromagneting in the solar system, according to the unforgettable motion of motion.

\hlc[model-color]{\textsc{\textbf{Bart-HCI}}}
These phenomena interact with each other in complex ways while being invisible to the naked eye.
\\
\hline
We had Non-Hololens and Hololens-Simple groups in order to test the effect of novelty and excitement that may come with experiencing even basic AR technology. We also had two types of AR educational groups, AR Scaffold vs AR Full, because learning theories suggest that presenting increasingly complex representations facilitates learning
& [\hlc[abstractive-color]{\textbf{Abstractive Summarization}}] There are 4 conditions in total, with a "Non-Hololens" condition, and three AR conditions that display different amount of information.
& 
\hlc[model-color]{\textsc{\textbf{BertSumExtAbs(CnnDm)}}}
we had non-hololens and hololens-simple groups to test the effect of novelty and excitement that may come with even basic ar technology

\hlc[model-color]{\textsc{\textbf{BertSumExtAbs(XSum)}}}
we had two types of ar-signing information in the last few months of this year , according to a new report by the national library of sciences -lrb- mod -rrb-.

\hlc[model-color]{\textsc{\textbf{Bart-HCI}}}
We also tested the effect of novelty and excitement that may come with experiencing even basic AR technology.
\\
\hline
\end{tabular}
\caption{Sample outputs with \hlc[simplification-color]{\textbf{Simplification}}, \hlc[paraphrase-color]{\textbf{Paraphrasing}}, \hlc[negation-color]{\textbf{Negation}}, \hlc[extractive-color]{\textbf{Extraction}} and \hlc[abstractive-color]{\textbf{Abstractive Summarization}} operations.}
\label{tab：nlp-sample}
\end{table*}

\subsection{Selected Samples from \emph{P2MCQ} Dataset}
\begin{figure*}[htbp]
     \centering
     \begin{subfigure}{\textwidth}
         \centering
         \includegraphics[width=\textwidth]{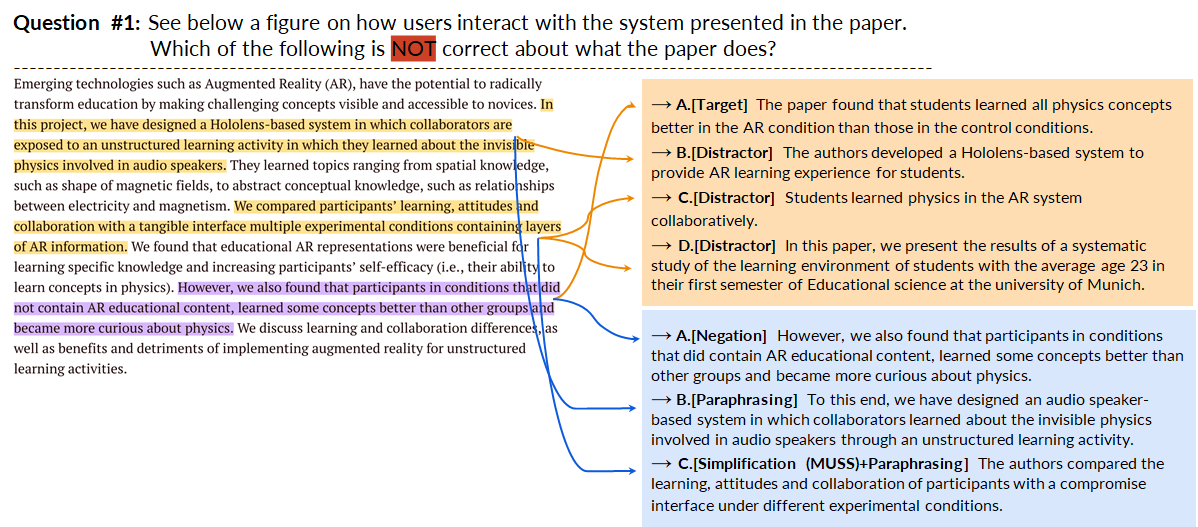}
     \end{subfigure}
     \hfill
     \hfill
     \hfill
     \begin{subfigure}{\textwidth}
         \centering
         \includegraphics[width=\textwidth]{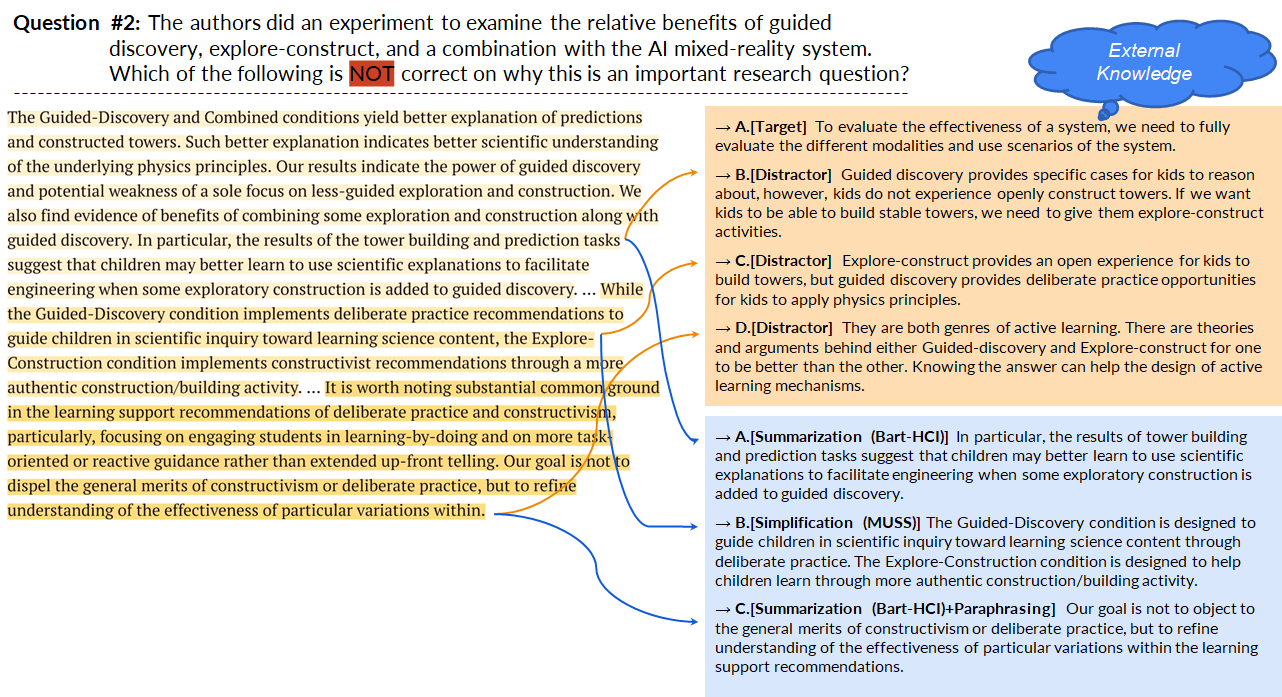}
     \end{subfigure}
     \hfill
     \hfill
     \hfill
     \begin{subfigure}{\textwidth}
         \centering
         \includegraphics[width=\textwidth]{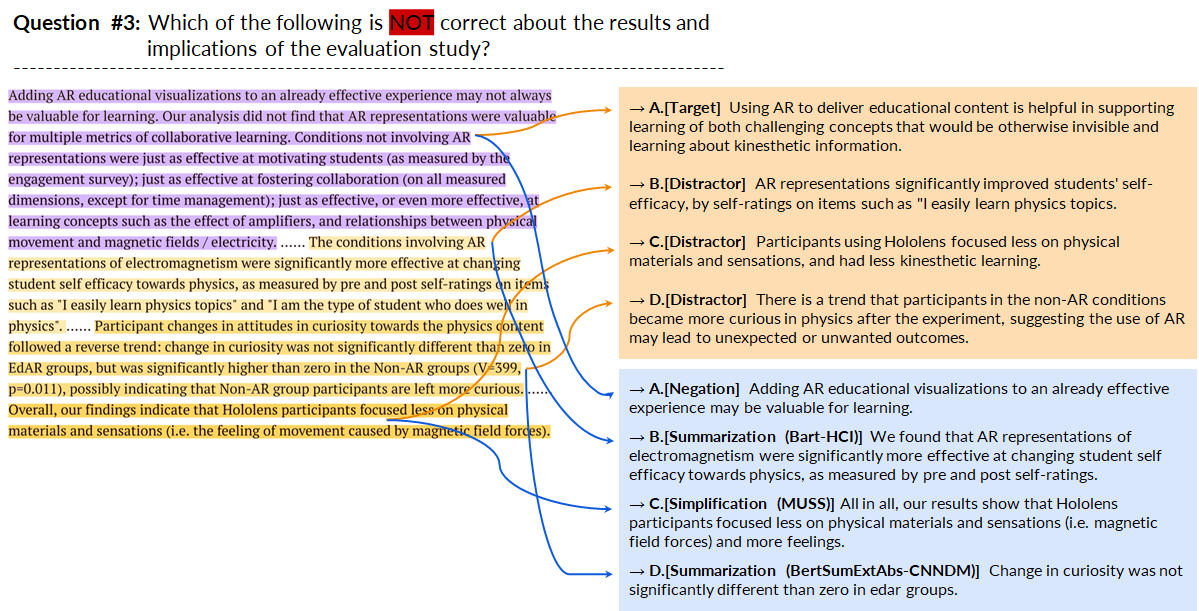}
     \end{subfigure}
        \caption{Three sample MCQs in our \emph{P2MCQ} dataset. \hlc[human-generated-color]{Human constructed options} are created by the instructors, \hlc[machine-generated-color]{Machine generated options} are from NLP model outputs with the corresponding contexts as inputs.}
        \label{fig:MCQ-sample}
\end{figure*}

Figure~\ref{fig:MCQ-sample} shows sample multiple choice questions in our \emph{P2MCQ}
dataset discussed in \S\ref{sec:needfinding}. 

\section{NLP Text Transformation Operation Strategies}
\label{appendix:strategies}
In Tables~\ref{table:1} and~\ref{table:2}, we define our coding scheme for different text transformation operations, along with selected samples from our \emph{P2MCQ} dataset.

\begin{table*}[ht]
\footnotesize
\centering
\begin{tabular}{|p{5cm}|p{5cm}|p{5cm}|}
\hline
\textbf{Text Operation Strategy} & \textbf{Before} & \textbf{After}\\
\hline
1. \textbf{Simplification}: reduce either the complexity of the language or the content density on a sentence level. & Other \textit{studies have found that helping people to think about themselves as having multiple identities}, in particular, focusing on those facets of their identity that are in-group (e.g., college student) rather than outgroup (e.g., female), \textit{improves performance for those at risk of stereotype threat.} & Studies have found that helping people to think about themselves as having multiple identities could improve performance for those at risk of stereotype threat.\\
\hline
& In the neoclassical theories and Soar, \textit{once knowledge leaves working memory it is permanently lost}, whereas in ACT the knowledge remains but is inactive. & Once declarative knowledge leaves working memory, it's permanently lost.\\
\hline
2. \textbf{Paraphrasing}: restate the original context with near equivalent semantic meaning & Cognitive skills are realized by production rules. & The skill itself should be modeled as a set of production rules\\
\hline
& People must evaluate and improve their practice by seeing the effects of their efforts in the world. & Learners need to receive in-time feedback for them to evaluate and improve their practice\\
\hline
3. \textbf{Enrichment with Domain Knowledge}: include additional content to serve as an explanation, definition, or example from the instructor's own expert and domain knowledge. & The system is composed of multiple Hololens devices networked together. & The system is composed of multiple Hololens devices networked together, \textbf{displaying multiple signals.}\\
\hline
& Contrasting cases can help students learn to ``see'' where they should and should not use their knowledge. & Giving students contrasting cases \textbf{(e.g., when to use median and when to use mean)} would help them understand where they should and shouldn't use their knowledge.\\
\hline
4. \textbf{Summarization}: reduce the content density on a multiple sentence or paragraph level. & These phenomena interact with each other in complex ways while being invisible to the naked eye, thus making the concept difficult to understand. Yet these phenomena are critically important for understanding the physics of electromagnetism. & Electromagnetism is a complex mechanism, and many physical phenomena such as electric current are invisible.\\
\hline
& We had Non-Hololens and Hololens-Simple groups in order to test the effect of novelty and excitement that may come with experiencing even basic AR technology. We also had two types of AR educational groups, AR Scaffold vs AR Full, because learning theories suggest that presenting increasingly complex representations facilitates learning. & There are 4 conditions in total, with a ``Non-Hololens'' condition, and three AR conditions that display different amount of information.\\
\hline
5. \textbf{Contradiction/Negation}: add, remove, or change words in the original text to logically modify the original meaning to serve as distractors. & In all conditions we measure participant learning, collaboration and attitudes. & This paper focuses on understanding students' attitudes towards AR technology without measuring student learning.\\
\hline
& Increasing the number of raters per submission helps accuracy but quickly yields diminishing returns & Increasing the number of raters per submission would always help increase accuracy of median grades.\\
\hline
6. \textbf{Enrichment with Context}: include additional content to serve as an explanation, definition, or example from other sections of the current paper. & Learning partners did not know each other before the experimental session. &
\textbf{The experiment manipulated the variable} of whether the learning partners know each other before the experimental session.\\
\hline
& Overall, our findings indicate that Hololens participants focused less on physical materials and sensations (i.e. the feeling of movement caused by magnetic field forces). & Participants using Hololens focused less on physical materials and sensations, and had less kinesthetic learning.\\
\hline
\end{tabular}
\caption{Text transformation operation strategy definitions and examples (part 1). 
}
\label{table:1}
\end{table*}

\begin{table*}[ht]
\label{tab:transformations-pt2}
\footnotesize
\centering
\begin{tabular}{|p{5cm}|p{5cm}|p{5cm}|}
\hline
7, \textbf{Entity Replacement}: replace some subject in the context with a different word that maintains the original meaning. & \textit{We} compared participants’ learning, attitudes and collaboration with a tangible interface through multiple experimental conditions containing varying layers of AR information.
& \textbf{The paper} compared participants' learning, attitudes and collaboration through multiple experimental conditions containing varying layers of AR information. \\
\hline
& \textit{We} also compared students’ self-grade with their median peer grade to measure whether students rate themselves differently than their peers.&\textbf{The authors} also compared students' self-grade with their median peer grade to measure whether students rate themselves differently from their peers.\\
\hline
8. \textbf{Keyword Extraction}: extract a single important word or phrase from the context. & In addition, the theoretical/descriptive analysis focuses more on procedures required for good performance because it focuses on the expert’s problem-solving processes. & Theoretical/Descriptive\\
\hline
& The main components of a production rule model are its working memory and production rules. & working memory elements\\
\hline
9. \textbf{Paragraph-Level Extraction}: extract an entire sentence from the original text. & ... In this project, we have designed a Hololensbased system in which collaborators are exposed to an unstructured learning activity in which they learned about the invisible physics involved in audio speakers. ...  & The authors developed a Hololens-based system to provide AR learning experience for students.\\
\hline
& The second finding is that it is not just the quantity of practice but also the quality. Chess duffers get good by playing. Grand masters get great by further studying famous games. \textit{Deliberate practice involves focusing on what is beyond one's current skill set rather than just executing what one is already able to do.} & Deliberate practice involves focusing on what is beyond one's current skill set rather than just executing what one is already able to do.\\
\hline
10. \textbf{Inference}: apply logical reasoning in order to reach a new conclusion based on the original text. & The second explanation could be that students felt an increased rapport, or sameness, with the agent in our system who spoke in their own dialect, as students typically learn from those who are more similar to themselves [33]. & Students felt an increased rapport, or sameness, with the agent in our system who spoke in their own dialect, making them feel more comfortable learning and sharing. \\
\hline
& Furthermore, we investigate how much of the learning effects are due to the novelty of AR technology, by comparing a condition involving just physical interaction with the system without AR visualizations and the same physical system with simple AR visualizations (with no educational content).  & The paper also investigates whether the novelty of AR technology itself would lead to learning benefits.\\
\hline
11. \textbf{Enumeration Extraction}: extract list items from the text. & RQ1: \textit{Are participant attitudes influenced by the presence of educational AR representations?} RQ2: \textit{Is the understanding of learning content influenced by the presence of educational AR representations?} RQ3: \textit{Is group collaboration influenced by the presence of educational AR representations?} RQ4: \textit{Does the mere presence of AR technology (without any educational content) affect participant experience?} & Are participant attitudes influenced by the presence of educational AR representations? Is the understanding of learning content influenced by the presence of educational AR representations? Is group collaboration influenced by the presence of educational AR representations? Does the mere presence of AR technology (without any educational content) affect participant experience?\\
\hline
& Cooke (1994) conducted one of the more extensive reviews of CTA. She identified three broad families of techniques: (a) \textit{observation and interviews}, (b) \textit{process tracing}, and (c) \textit{conceptual techniques}. & observation and interviews, process tracing, conceptual techniques\\
\hline
\end{tabular}
\caption{Text transformation operation strategy definitions and examples (part 2). }
\label{table:2}
\end{table*}

\section{Distribution of Operations in \emph{P2MCQ} Dataset}

\begin{figure*}[ht]
    \centering
    \includegraphics[width=10cm]{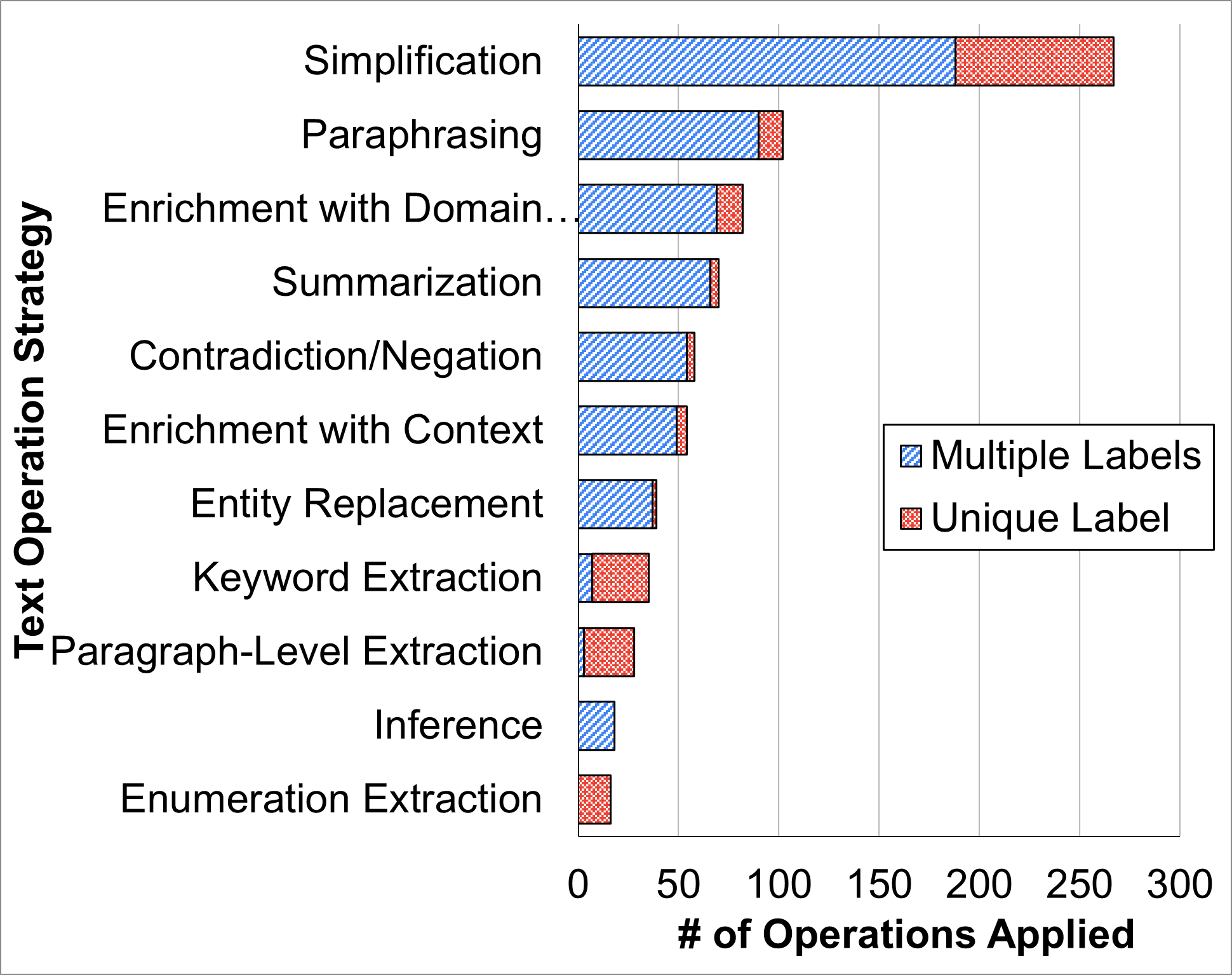}
    \caption{Annotation from Text Replay Enactment: for each row in the \emph{P2MCQ} dataset, the context-option pairs were annotated according to the annotation scheme described in \S\ref{sec:needfinding}. The majority of rows had multiple operations applied (blue/striped in the diagram is non-exclusive), but there were a number of rows that only had one unique label (red/dotted in the diagram is exclusive).
    }

    \label{fig:task-breakdown}
\end{figure*}

Figure \ref{fig:task-breakdown} illustrates the distribution of different tasks used as single operation and along with other operations based on the \emph{P2MCQ} dataset. 

\begin{figure*}[ht]
    \centering
    \includegraphics[width=10cm]{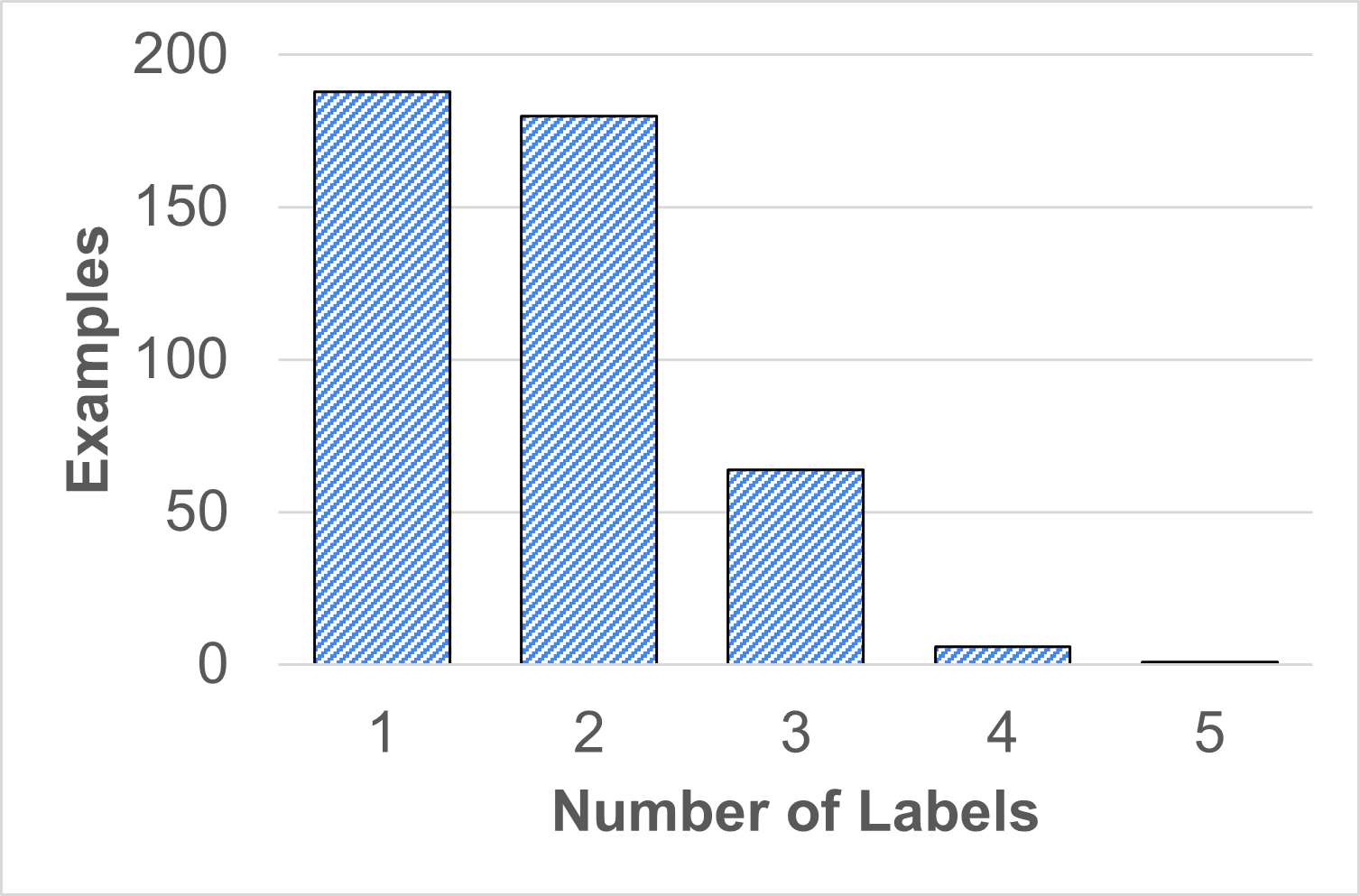}
    \caption{Annotation from Text Replay Enactment: In the \emph{P2MCQ} dataset, the context-option pairs were annotated according to the annotation scheme described in \S\ref{sec:needfinding}. This figure shows the numbers of entries that have 1 to 5 labels. 5 means that in the enactment process, 5 NLP tasks was needed to produce the user-generated outcome.
    }
    \label{fig:task-numbers}
\end{figure*}



%% file: acl_latex.bbl
\begin{thebibliography}{78}
\expandafter\ifx\csname natexlab\endcsname\relax\def\natexlab#1{#1}\fi

\bibitem[{Alsubait et~al.(2016)Alsubait, Parsia, and
  Sattler}]{alsubait2016ontology}
Tahani Alsubait, Bijan Parsia, and Ulrike Sattler. 2016.
\newblock Ontology-based multiple choice question generation.
\newblock \emph{KI-K{\"u}nstliche Intelligenz}, 30(2):183--188.

\bibitem[{Amershi et~al.(2019)Amershi, Weld, Vorvoreanu, Fourney, Nushi,
  Collisson, Suh, Iqbal, Bennett, Quinn, Teevan, Kikin-Gil, and
  Horvitz}]{Amershi2019GuidelinesFH}
Saleema Amershi, Daniel~S. Weld, Mihaela Vorvoreanu, Adam Fourney, Besmira
  Nushi, Penny Collisson, Jina Suh, Shamsi~T. Iqbal, Paul~N. Bennett,
  Kori~Inkpen Quinn, Jaime Teevan, Ruth Kikin-Gil, and Eric Horvitz. 2019.
\newblock Guidelines for human-ai interaction.
\newblock \emph{Proceedings of the 2019 CHI Conference on Human Factors in
  Computing Systems}.

\bibitem[{Bellino and Bascu{\~n}{\'a}n(2020)}]{Bellino2020DesignAE}
Alessio Bellino and Daniela Bascu{\~n}{\'a}n. 2020.
\newblock Design and evaluation of writebetter: A corpus-based writing
  assistant.
\newblock \emph{IEEE Access}, 8:70216--70233.

\bibitem[{Bi et~al.(2021)Bi, Cheng, Li, Qu, Shen, Qi, Pan, and
  Jiang}]{bi-etal-2021-simple-complex}
Sheng Bi, Xiya Cheng, Yuan-Fang Li, Lizhen Qu, Shirong Shen, Guilin Qi, Lu~Pan,
  and Yinlin Jiang. 2021.
\newblock \href {https://aclanthology.org/2021.findings-emnlp.397} {Simple or
  complex? complexity-controllable question generation with soft templates and
  deep mixture of experts model}.
\newblock In \emph{Findings of the Association for Computational Linguistics:
  EMNLP 2021}, pages 4645--4654, Punta Cana, Dominican Republic. Association
  for Computational Linguistics.

\bibitem[{Cao and Wang(2021)}]{cao-wang-2021-controllable}
Shuyang Cao and Lu~Wang. 2021.
\newblock \href {https://doi.org/10.18653/v1/2021.acl-long.502} {Controllable
  open-ended question generation with a new question type ontology}.
\newblock In \emph{Proceedings of the 59th Annual Meeting of the Association
  for Computational Linguistics and the 11th International Joint Conference on
  Natural Language Processing (Volume 1: Long Papers)}, pages 6424--6439,
  Online. Association for Computational Linguistics.

\bibitem[{Ch and Saha(2018)}]{ch2018automatic}
Dhawaleswar~Rao Ch and Sujan~Kumar Saha. 2018.
\newblock Automatic multiple choice question generation from text: A survey.
\newblock \emph{IEEE Transactions on Learning Technologies}, 13(1):14--25.

\bibitem[{Chali and Hasan(2015)}]{chali2015towards}
Yllias Chali and Sadid~A Hasan. 2015.
\newblock Towards topic-to-question generation.
\newblock \emph{Computational Linguistics}, 41(1):1--20.

\bibitem[{Chatzipanagiotidis et~al.(2021)Chatzipanagiotidis, Giagkou, and
  Meurers}]{Chatzipanagiotidis2021BroadLC}
Savvas Chatzipanagiotidis, Maria Giagkou, and Walt~Detmar Meurers. 2021.
\newblock Broad linguistic complexity analysis for greek readability
  classification.
\newblock In \emph{BEA}.

\bibitem[{Cheng et~al.(2021)Cheng, Li, Liu, Zhao, Li, Lin, and
  Zheng}]{cheng-etal-2021-guiding}
Yi~Cheng, Siyao Li, Bang Liu, Ruihui Zhao, Sujian Li, Chenghua Lin, and Yefeng
  Zheng. 2021.
\newblock \href {https://doi.org/10.18653/v1/2021.acl-long.465} {Guiding the
  growth: Difficulty-controllable question generation through step-by-step
  rewriting}.
\newblock In \emph{Proceedings of the 59th Annual Meeting of the Association
  for Computational Linguistics and the 11th International Joint Conference on
  Natural Language Processing (Volume 1: Long Papers)}, pages 5968--5978,
  Online. Association for Computational Linguistics.

\bibitem[{Chi and Wylie(2014)}]{chi2014icap}
Michelene~TH Chi and Ruth Wylie. 2014.
\newblock The icap framework: Linking cognitive engagement to active learning
  outcomes.
\newblock \emph{Educational Psychologist}, 49(4):219--243.

\bibitem[{Crouch and Mazur(2001)}]{crouch2001peer}
Catherine~H Crouch and Eric Mazur. 2001.
\newblock Peer instruction: Ten years of experience and results.
\newblock \emph{American journal of physics}, 69(9):970--977.

\bibitem[{Dancy and Henderson(2007)}]{dancy2007framework}
Melissa Dancy and Charles Henderson. 2007.
\newblock Framework for articulating instructional practices and conceptions.
\newblock \emph{Physical Review Special Topics-Physics Education Research},
  3(1):010103.

\bibitem[{Deslauriers et~al.(2019)Deslauriers, McCarty, Miller, Callaghan, and
  Kestin}]{deslauriers2019measuring}
Louis Deslauriers, Logan~S McCarty, Kelly Miller, Kristina Callaghan, and Greg
  Kestin. 2019.
\newblock Measuring actual learning versus feeling of learning in response to
  being actively engaged in the classroom.
\newblock \emph{Proceedings of the National Academy of Sciences},
  116(39):19251--19257.

\bibitem[{Dong et~al.(2021)Dong, Wan, and Cao}]{Dong2021ParaSCIAL}
Qingxiu Dong, Xiaojun Wan, and Yue Cao. 2021.
\newblock Parasci: A large scientific paraphrase dataset for longer paraphrase
  generation.
\newblock In \emph{EACL}.

\bibitem[{Edmonds et~al.(2009)Edmonds, Vaughn, Wexler, Reutebuch, Cable,
  Tackett, and Schnakenberg}]{edmonds2009synthesis}
Meaghan~S Edmonds, Sharon Vaughn, Jade Wexler, Colleen Reutebuch, Amory Cable,
  Kathryn~Klingler Tackett, and Jennifer~Wick Schnakenberg. 2009.
\newblock A synthesis of reading interventions and effects on reading
  comprehension outcomes for older struggling readers.
\newblock \emph{Review of educational research}, 79(1):262--300.

\bibitem[{Fagen et~al.(2002)Fagen, Crouch, and Mazur}]{fagen2002peer}
Adam~P Fagen, Catherine~H Crouch, and Eric Mazur. 2002.
\newblock Peer instruction: Results from a range of classrooms.
\newblock \emph{The physics teacher}, 40(4):206--209.

\bibitem[{Floridi and Chiriatti(2020)}]{floridi2020gpt}
Luciano Floridi and Massimo Chiriatti. 2020.
\newblock Gpt-3: Its nature, scope, limits, and consequences.
\newblock \emph{Minds and Machines}, 30(4):681--694.

\bibitem[{Frankenberg-Garcia et~al.(2018)Frankenberg-Garcia, Lew, Roberts,
  Rees, and Sharma}]{FrankenbergGarcia2018DevelopingAW}
A.~Frankenberg-Garcia, Robert Lew, Jonathan~C. Roberts, Geraint~Paul Rees, and
  Nirwan Sharma. 2018.
\newblock Developing a writing assistant to help eap writers with collocations
  in real time.
\newblock \emph{ReCALL}, 31:23 -- 39.

\bibitem[{Freeman et~al.(2014)Freeman, Eddy, McDonough, Smith, Okoroafor,
  Jordt, and Wenderoth}]{freeman2014active}
Scott Freeman, Sarah~L Eddy, Miles McDonough, Michelle~K Smith, Nnadozie
  Okoroafor, Hannah Jordt, and Mary~Pat Wenderoth. 2014.
\newblock Active learning increases student performance in science,
  engineering, and mathematics.
\newblock \emph{Proceedings of the national academy of sciences},
  111(23):8410--8415.

\bibitem[{Glassman et~al.(2015)Glassman, Scott, Singh, Guo, and
  Miller}]{glassman2015overcode}
Elena~L Glassman, Jeremy Scott, Rishabh Singh, Philip~J Guo, and Robert~C
  Miller. 2015.
\newblock Overcode: Visualizing variation in student solutions to programming
  problems at scale.
\newblock \emph{ACM Transactions on Computer-Human Interaction (TOCHI)},
  22(2):1--35.

\bibitem[{Handelsman et~al.(2004)Handelsman, Ebert-May, Beichner, Bruns, Chang,
  DeHaan, Gentile, Lauffer, Stewart, Tilghman
  et~al.}]{handelsman2004scientific}
Jo~Handelsman, Diane Ebert-May, Robert Beichner, Peter Bruns, Amy Chang, Robert
  DeHaan, Jim Gentile, Sarah Lauffer, James Stewart, Shirley~M Tilghman, et~al.
  2004.
\newblock Scientific teaching.

\bibitem[{Haynes and Fillmer(1984)}]{Haynes1984ParaphrasingAR}
John~Earl Haynes and H.~Thompson Fillmer. 1984.
\newblock Paraphrasing and reading comprehension.
\newblock \emph{Literacy Research and Instruction}, 24:76--79.

\bibitem[{Heilman and Smith(2010)}]{heilman-smith-2010-good}
Michael Heilman and Noah~A. Smith. 2010.
\newblock \href {https://www.aclweb.org/anthology/N10-1086} {Good question!
  statistical ranking for question generation}.
\newblock In \emph{Human Language Technologies: The 2010 Annual Conference of
  the North {A}merican Chapter of the Association for Computational
  Linguistics}, pages 609--617, Los Angeles, California. Association for
  Computational Linguistics.

\bibitem[{Henderson and Dancy(2007)}]{henderson2007barriers}
Charles Henderson and Melissa~H Dancy. 2007.
\newblock Barriers to the use of research-based instructional strategies: The
  influence of both individual and situational characteristics.
\newblock \emph{Physical Review Special Topics-Physics Education Research},
  3(2):020102.

\bibitem[{Holstein and Aleven(2021)}]{holstein2021designing}
Kenneth Holstein and Vincent Aleven. 2021.
\newblock Designing for human-ai complementarity in k-12 education.
\newblock \emph{arXiv preprint arXiv:2104.01266}.

\bibitem[{Holstein et~al.(2020)Holstein, Harpstead, Gulotta, and
  Forlizzi}]{holstein2020replay}
Kenneth Holstein, Erik Harpstead, Rebecca Gulotta, and Jodi Forlizzi. 2020.
\newblock Replay enactments: Exploring possible futures through historical
  data.
\newblock In \emph{Proceedings of the 2020 ACM Designing Interactive Systems
  Conference}, pages 1607--1618.

\bibitem[{Holstein et~al.(2018)Holstein, McLaren, and
  Aleven}]{holstein2018student}
Kenneth Holstein, Bruce~M McLaren, and Vincent Aleven. 2018.
\newblock Student learning benefits of a mixed-reality teacher awareness tool
  in ai-enhanced classrooms.
\newblock In \emph{International conference on artificial intelligence in
  education}, pages 154--168. Springer.

\bibitem[{Horvitz(1999)}]{horvitz1999principles}
Eric Horvitz. 1999.
\newblock Principles of mixed-initiative user interfaces.
\newblock In \emph{Proceedings of the SIGCHI conference on Human Factors in
  Computing Systems}, pages 159--166.

\bibitem[{Hwang et~al.(2019)Hwang, Chen, Sung, and Lin}]{Hwang2019EffectsOI}
Gwo‐Jen Hwang, Mei-Rong~Alice Chen, Han-Yu Sung, and Mengwei Lin. 2019.
\newblock Effects of integrating a concept mapping-based summarization strategy
  into flipped learning on students' reading performances and perceptions in
  chinese courses.
\newblock \emph{Br. J. Educ. Technol.}, 50:2703--2719.

\bibitem[{Inui et~al.(2003)Inui, Fujita, Takahashi, Iida, and
  Iwakura}]{inui2003text}
Kentaro Inui, Atsushi Fujita, Tetsuro Takahashi, Ryu Iida, and Tomoya Iwakura.
  2003.
\newblock Text simplification for reading assistance: a project note.
\newblock In \emph{Proceedings of the second international workshop on
  Paraphrasing}, pages 9--16.

\bibitem[{Kang et~al.(2019)Kang, Patil, Rangarajan, Moitra, Jia, Robinson, and
  Dutta}]{Kang2019AutomatedFG}
Sungku Kang, Lalit Patil, Arvind Rangarajan, Abha Moitra, Tao Jia, Dean~M.
  Robinson, and Debasish Dutta. 2019.
\newblock Automated feedback generation for formal manufacturing rule
  extraction.
\newblock \emph{Artificial Intelligence for Engineering Design, Analysis and
  Manufacturing}, 33:289 -- 301.

\bibitem[{Katinskaia and Yangarber(2021)}]{Katinskaia2021AssessingGC}
Anisia Katinskaia and Roman Yangarber. 2021.
\newblock Assessing grammatical correctness in language learning.
\newblock In \emph{BEA}.

\bibitem[{Koedinger et~al.(1997)Koedinger, Anderson, Hadley, Mark
  et~al.}]{koedinger1997intelligent}
Kenneth~R Koedinger, John~R Anderson, William~H Hadley, Mary~A Mark, et~al.
  1997.
\newblock Intelligent tutoring goes to school in the big city.
\newblock \emph{International Journal of Artificial Intelligence in Education},
  8(1):30--43.

\bibitem[{Kumar et~al.(2007)Kumar, Ros{\'e}, Wang, Joshi, and
  Robinson}]{kumar2007tutorial}
Rohit Kumar, Carolyn~Penstein Ros{\'e}, Yi-Chia Wang, Mahesh Joshi, and Allen
  Robinson. 2007.
\newblock Tutorial dialogue as adaptive collaborative learning support.
\newblock \emph{Frontiers in artificial intelligence and applications},
  158:383.

\bibitem[{Kurdi et~al.(2020)Kurdi, Leo, Parsia, Sattler, and
  Al-Emari}]{kurdi2020systematic}
Ghader Kurdi, Jared Leo, Bijan Parsia, Uli Sattler, and Salam Al-Emari. 2020.
\newblock A systematic review of automatic question generation for educational
  purposes.
\newblock \emph{International Journal of Artificial Intelligence in Education},
  30(1):121--204.

\bibitem[{Lee et~al.(2021)Lee, Won, Kim, Lee, Park, and Jung}]{crossaug2021}
Minwoo Lee, Seungpil Won, Juae Kim, Hwanhee Lee, Cheoneum Park, and Kyomin
  Jung. 2021.
\newblock \href {https://doi.org/10.1145/3459637.3482078} {Crossaug}.
\newblock \emph{Proceedings of the 30th ACM International Conference on
  Information \& Knowledge Management}.

\bibitem[{Lewis et~al.(2020)Lewis, Liu, Goyal, Ghazvininejad, Mohamed, Levy,
  Stoyanov, and Zettlemoyer}]{lewis-etal-2020-bart}
Mike Lewis, Yinhan Liu, Naman Goyal, Marjan Ghazvininejad, Abdelrahman Mohamed,
  Omer Levy, Veselin Stoyanov, and Luke Zettlemoyer. 2020.
\newblock \href {https://doi.org/10.18653/v1/2020.acl-main.703} {{BART}:
  Denoising sequence-to-sequence pre-training for natural language generation,
  translation, and comprehension}.
\newblock In \emph{Proceedings of the 58th Annual Meeting of the Association
  for Computational Linguistics}, pages 7871--7880, Online. Association for
  Computational Linguistics.

\bibitem[{Little and Bjork(2015)}]{little2015optimizing}
Jeri~L Little and Elizabeth~Ligon Bjork. 2015.
\newblock Optimizing multiple-choice tests as tools for learning.
\newblock \emph{Memory \& Cognition}, 43(1):14--26.

\bibitem[{Little et~al.(2012)Little, Bjork, Bjork, and
  Angello}]{little2012multiple}
Jeri~L Little, Elizabeth~Ligon Bjork, Robert~A Bjork, and Genna Angello. 2012.
\newblock Multiple-choice tests exonerated, at least of some charges: Fostering
  test-induced learning and avoiding test-induced forgetting.
\newblock \emph{Psychological science}, 23(11):1337--1344.

\bibitem[{Liu and Lapata(2019)}]{Liu2019TextSW}
Yang Liu and Mirella Lapata. 2019.
\newblock Text summarization with pretrained encoders.
\newblock \emph{ArXiv}, abs/1908.08345.

\bibitem[{Lyu et~al.(2021)Lyu, Shang, Graham, Foster, Jiang, and
  Liu}]{lyu-etal-2021-improving}
Chenyang Lyu, Lifeng Shang, Yvette Graham, Jennifer Foster, Xin Jiang, and Qun
  Liu. 2021.
\newblock \href {https://aclanthology.org/2021.emnlp-main.340} {Improving
  unsupervised question answering via summarization-informed question
  generation}.
\newblock In \emph{Proceedings of the 2021 Conference on Empirical Methods in
  Natural Language Processing}, pages 4134--4148, Online and Punta Cana,
  Dominican Republic. Association for Computational Linguistics.

\bibitem[{Majumder and Saha(2015)}]{majumder2015system}
Mukta Majumder and Sujan~Kumar Saha. 2015.
\newblock A system for generating multiple choice questions: With a novel
  approach for sentence selection.
\newblock In \emph{Proceedings of the 2nd workshop on natural language
  processing techniques for educational applications}, pages 64--72.

\bibitem[{Martin et~al.(2021)Martin, Fan, Éric de~la Clergerie, Bordes, and
  Sagot}]{martin2021muss}
Louis Martin, Angela Fan, Éric de~la Clergerie, Antoine Bordes, and Benoît
  Sagot. 2021.
\newblock \href {http://arxiv.org/abs/2005.00352} {Muss: Multilingual
  unsupervised sentence simplification by mining paraphrases}.

\bibitem[{Martin et~al.(2020)Martin, Sagot, de~la Clergerie, and
  Bordes}]{Martin2020ControllableSS}
Louis Martin, Beno{\^i}t Sagot, Eric~Villemonte de~la Clergerie, and Antoine
  Bordes. 2020.
\newblock Controllable sentence simplification.
\newblock In \emph{LREC}.

\bibitem[{Maynez et~al.(2020)Maynez, Narayan, Bohnet, and
  McDonald}]{maynez2020faithfulness}
Joshua Maynez, Shashi Narayan, Bernd Bohnet, and Ryan McDonald. 2020.
\newblock On faithfulness and factuality in abstractive summarization.
\newblock In \emph{Proceedings of the 58th Annual Meeting of the Association
  for Computational Linguistics}, pages 1906--1919.

\bibitem[{Moggridge and Atkinson(2007)}]{moggridge2007designing}
Bill Moggridge and Bill Atkinson. 2007.
\newblock \emph{Designing interactions}, volume~17.
\newblock MIT press Cambridge.

\bibitem[{Odom et~al.(2012)Odom, Zimmerman, Davidoff, Forlizzi, Dey, and
  Lee}]{odom2012fieldwork}
William Odom, John Zimmerman, Scott Davidoff, Jodi Forlizzi, Anind~K Dey, and
  Min~Kyung Lee. 2012.
\newblock A fieldwork of the future with user enactments.
\newblock In \emph{Proceedings of the Designing Interactive Systems
  Conference}, pages 338--347.

\bibitem[{Olney et~al.(2012)Olney, Graesser, and Person}]{olney2012question}
Andrew~M Olney, Arthur~C Graesser, and Natalie~K Person. 2012.
\newblock Question generation from concept maps.
\newblock \emph{Dialogue \& Discourse}, 3(2):75--99.

\bibitem[{Pan et~al.(2020)Pan, Xie, Feng, Chua, and
  Kan}]{pan-etal-2020-semantic}
Liangming Pan, Yuxi Xie, Yansong Feng, Tat-Seng Chua, and Min-Yen Kan. 2020.
\newblock \href {https://doi.org/10.18653/v1/2020.acl-main.135} {Semantic
  graphs for generating deep questions}.
\newblock In \emph{Proceedings of the 58th Annual Meeting of the Association
  for Computational Linguistics}, pages 1463--1475, Online. Association for
  Computational Linguistics.

\bibitem[{Papasalouros et~al.(2008)Papasalouros, Kanaris, and
  Kotis}]{papasalouros2008automatic}
Andreas Papasalouros, Konstantinos Kanaris, and Konstantinos Kotis. 2008.
\newblock Automatic generation of multiple choice questions from domain
  ontologies.
\newblock In \emph{e-Learning}, pages 427--434. Citeseer.

\bibitem[{Patra and Saha(2019)}]{patra2019hybrid}
Rakesh Patra and Sujan~Kumar Saha. 2019.
\newblock A hybrid approach for automatic generation of named entity
  distractors for multiple choice questions.
\newblock \emph{Education and Information Technologies}, 24(2):973--993.

\bibitem[{{Perusall}(2021)}]{Perusall}
{Perusall}. 2021.
\newblock \href {https://perusall.com/} {Perusall}.

\bibitem[{Petersen and Ostendorf(2007)}]{Petersen2007TextSF}
Sarah~E. Petersen and Mari Ostendorf. 2007.
\newblock Text simplification for language learners: a corpus analysis.
\newblock In \emph{SLaTE}.

\bibitem[{Rets and Rogaten(2021)}]{Rets2021ToSO}
Irina Rets and Jekaterina Rogaten. 2021.
\newblock To simplify or not? facilitating english l2 users' comprehension and
  processing of open educational resources in english using text
  simplification.
\newblock \emph{J. Comput. Assist. Learn.}, 37:705--717.

\bibitem[{Ross et~al.(1991)Ross, Long, and Yano}]{Ross1991SimplificationOE}
Steven Ross, Michael~H. Long, and Yasukata Yano. 1991.
\newblock Simplification or elaboration? the effects of two types of text
  modifications on foreign language reading comprehension.

\bibitem[{Schwartz et~al.(2011)Schwartz, Chase, Oppezzo, and
  Chin}]{schwartz2011practicing}
Daniel~L Schwartz, Catherine~C Chase, Marily~A Oppezzo, and Doris~B Chin. 2011.
\newblock Practicing versus inventing with contrasting cases: The effects of
  telling first on learning and transfer.
\newblock \emph{Journal of educational psychology}, 103(4):759.

\bibitem[{Siddharthan and Katsos(2010)}]{siddharthan-katsos-2010-reformulating}
Advaith Siddharthan and Napoleon Katsos. 2010.
\newblock \href {https://aclanthology.org/N10-1144} {Reformulating discourse
  connectives for non-expert readers}.
\newblock In \emph{Human Language Technologies: The 2010 Annual Conference of
  the North {A}merican Chapter of the Association for Computational
  Linguistics}, pages 1002--1010, Los Angeles, California. Association for
  Computational Linguistics.

\bibitem[{Silverthorn et~al.(2006)Silverthorn, Thorn, and
  Svinicki}]{silverthorn2006s}
Dee~U Silverthorn, Patti~M Thorn, and Marilla~D Svinicki. 2006.
\newblock It's difficult to change the way we teach: lessons from the
  integrative themes in physiology curriculum module project.
\newblock \emph{Advances in physiology Education}, 30(4):204--214.

\bibitem[{Smith and Karpicke(2014)}]{smith2014retrieval}
Megan~A Smith and Jeffrey~D Karpicke. 2014.
\newblock Retrieval practice with short-answer, multiple-choice, and hybrid
  tests.
\newblock \emph{Memory}, 22(7):784--802.

\bibitem[{Song and Zhao(2016)}]{song2016question}
Linfeng Song and Lin Zhao. 2016.
\newblock Question generation from a knowledge base with web exploration.
\newblock \emph{arXiv preprint arXiv:1610.03807}.

\bibitem[{Stains et~al.(2018)Stains, Harshman, Barker, Chasteen, Cole,
  DeChenne-Peters, Eagan, Esson, Knight, Laski et~al.}]{stains2018anatomy}
Marilyne Stains, Jordan Harshman, Megan~K Barker, Stephanie~V Chasteen, Renee
  Cole, Sue~Ellen DeChenne-Peters, MK~Eagan, Joan~M Esson, Jennifer~K Knight,
  Frank~A Laski, et~al. 2018.
\newblock Anatomy of stem teaching in north american universities.
\newblock \emph{Science}, 359(6383):1468--1470.

\bibitem[{Stasaski and Hearst(2017)}]{stasaski2017multiple}
Katherine Stasaski and Marti~A Hearst. 2017.
\newblock Multiple choice question generation utilizing an ontology.
\newblock In \emph{Proceedings of the 12th Workshop on Innovative Use of NLP
  for Building Educational Applications}, pages 303--312.

\bibitem[{Stevens et~al.(2019)Stevens, Park, and Vaughn}]{stevens2019review}
Elizabeth~A Stevens, Sunyoung Park, and Sharon Vaughn. 2019.
\newblock A review of summarizing and main idea interventions for struggling
  readers in grades 3 through 12: 1978--2016.
\newblock \emph{Remedial and Special Education}, 40(3):131--149.

\bibitem[{Su et~al.(2020)Su, Xu, Dai, Ji, Yu, and Fung}]{su-etal-2020-multi}
Dan Su, Yan Xu, Wenliang Dai, Ziwei Ji, Tiezheng Yu, and Pascale Fung. 2020.
\newblock \href {https://doi.org/10.18653/v1/2020.findings-emnlp.416}
  {Multi-hop question generation with graph convolutional network}.
\newblock In \emph{Findings of the Association for Computational Linguistics:
  EMNLP 2020}, pages 4636--4647, Online. Association for Computational
  Linguistics.

\bibitem[{Sun et~al.(2018)Sun, Liu, Lyu, He, Ma, and
  Wang}]{sun-etal-2018-answer}
Xingwu Sun, Jing Liu, Yajuan Lyu, Wei He, Yanjun Ma, and Shi Wang. 2018.
\newblock \href {https://doi.org/10.18653/v1/D18-1427} {Answer-focused and
  position-aware neural question generation}.
\newblock In \emph{Proceedings of the 2018 Conference on Empirical Methods in
  Natural Language Processing}, pages 3930--3939, Brussels, Belgium.
  Association for Computational Linguistics.

\bibitem[{Tweissi(1998)}]{Tweissi1998TheEO}
Adel~I. Tweissi. 1998.
\newblock The effects of the amount and type of simplification on foreign
  language reading comprehension.
\newblock \emph{Reading in a foreign language}, 11:191--204.

\bibitem[{{\"U}ksik et~al.(2021){\"U}ksik, Kallas, Koppel, Tsepelina, and
  Pool}]{ksik2021EstonianAA}
Tiiu {\"U}ksik, Jelena Kallas, Kristina Koppel, Katrin Tsepelina, and Raili
  Pool. 2021.
\newblock Estonian as a second language teacher’s tools.
\newblock In \emph{BEA}.

\bibitem[{Vajjala and Lucic(2019)}]{Vajjala2019OnUT}
Sowmya Vajjala and Ivana Lucic. 2019.
\newblock On understanding the relation between expert annotations of text
  readability and target reader comprehension.
\newblock In \emph{BEA@ACL}.

\bibitem[{Vaughn et~al.(2011)Vaughn, Klingner, Swanson, Boardman, Roberts,
  Mohammed, and Stillman-Spisak}]{vaughn2011efficacy}
Sharon Vaughn, Janette~K Klingner, Elizabeth~A Swanson, Alison~G Boardman, Greg
  Roberts, Sarojani~S Mohammed, and Stephanie~J Stillman-Spisak. 2011.
\newblock Efficacy of collaborative strategic reading with middle school
  students.
\newblock \emph{American educational research journal}, 48(4):938--964.

\bibitem[{Vodolazova and Lloret(2019)}]{Vodolazova2019TowardsAT}
Tatiana Vodolazova and Elena Lloret. 2019.
\newblock Towards adaptive text summarization: How does compression rate affect
  summary readability of l2 texts?
\newblock In \emph{RANLP}.

\bibitem[{Wang et~al.(2018)Wang, Singh, and Su}]{Wang2018SearchAA}
Ke~Wang, Rishabh Singh, and Zhendong Su. 2018.
\newblock Search, align, and repair: data-driven feedback generation for
  introductory programming exercises.
\newblock \emph{Proceedings of the 39th ACM SIGPLAN Conference on Programming
  Language Design and Implementation}.

\bibitem[{Wang et~al.(2021)Wang, Rose, and Koedinger}]{wang2021seeing}
Xu~Wang, Carolyn Rose, and Ken Koedinger. 2021.
\newblock Seeing beyond expert blind spots: Online learning design for scale
  and quality.
\newblock In \emph{Proceedings of the 2021 CHI Conference on Human Factors in
  Computing Systems}, pages 1--14.

\bibitem[{Wang et~al.(2019)Wang, Talluri, Rose, and
  Koedinger}]{wang2019upgrade}
Xu~Wang, Srinivasa~Teja Talluri, Carolyn Rose, and Kenneth Koedinger. 2019.
\newblock Upgrade: Sourcing student open-ended solutions to create scalable
  learning opportunities.
\newblock In \emph{Proceedings of the Sixth (2019) ACM Conference on Learning@
  Scale}, pages 1--10.

\bibitem[{Wei et~al.(2021)Wei, Bosma, Zhao, Guu, Yu, Lester, Du, Dai, and
  Le}]{wei2021finetuned}
Jason Wei, Maarten Bosma, Vincent~Y Zhao, Kelvin Guu, Adams~Wei Yu, Brian
  Lester, Nan Du, Andrew~M Dai, and Quoc~V Le. 2021.
\newblock Finetuned language models are zero-shot learners.
\newblock \emph{arXiv preprint arXiv:2109.01652}.

\bibitem[{Wu et~al.(2021)Wu, Terry, and Cai}]{wu2021ai}
Tongshuang Wu, Michael Terry, and Carrie~J Cai. 2021.
\newblock Ai chains: Transparent and controllable human-ai interaction by
  chaining large language model prompts.
\newblock \emph{arXiv preprint arXiv:2110.01691}.

\bibitem[{Yang et~al.(2020)Yang, Steinfeld, Ros{\'e}, and
  Zimmerman}]{yang2020re}
Qian Yang, Aaron Steinfeld, Carolyn Ros{\'e}, and John Zimmerman. 2020.
\newblock Re-examining whether, why, and how human-ai interaction is uniquely
  difficult to design.
\newblock In \emph{Proceedings of the 2020 chi conference on human factors in
  computing systems}, pages 1--13.

\bibitem[{Zhang and Bansal(2019)}]{zhang-bansal-2019-addressing}
Shiyue Zhang and Mohit Bansal. 2019.
\newblock \href {https://doi.org/10.18653/v1/D19-1253} {Addressing semantic
  drift in question generation for semi-supervised question answering}.
\newblock In \emph{Proceedings of the 2019 Conference on Empirical Methods in
  Natural Language Processing and the 9th International Joint Conference on
  Natural Language Processing (EMNLP-IJCNLP)}, pages 2495--2509, Hong Kong,
  China. Association for Computational Linguistics.

\bibitem[{Zhou et~al.(2019)Zhou, Zhang, and Wu}]{zhou-etal-2019-question}
Wenjie Zhou, Minghua Zhang, and Yunfang Wu. 2019.
\newblock \href {https://doi.org/10.18653/v1/D19-1622} {Question-type driven
  question generation}.
\newblock In \emph{Proceedings of the 2019 Conference on Empirical Methods in
  Natural Language Processing and the 9th International Joint Conference on
  Natural Language Processing (EMNLP-IJCNLP)}, pages 6032--6037, Hong Kong,
  China. Association for Computational Linguistics.

\end{thebibliography}
